 \documentclass[iop]{emulateapj}
 
\shorttitle{Field OB Star IMF}
\shortauthors{Lamb, J. B.,  et al.}

\usepackage{epsfig}

\begin{document}

\newcommand{\Mcl}{M_{\rm cl}}
\newcommand{\Mcllo}{M_{\rm cl,lo}}
\newcommand{\mlo}{m_{\rm lo}}
\newcommand{\mhi}{m_{\rm hi}}
\newcommand{\mup}{m_{\rm up}}
\newcommand{\Nlo}{N_{*,{\rm lo}}}
\newcommand{\mmax}{m_{\rm max}}
\newcommand{\mmaxtwo}{m_{\rm max,2}}
\newcommand{\mratio}{m_{\rm max,2} / m_{\rm max}}

\slugcomment{Resubmitted Dec 5, 2012}


\title{The Initial Mass Function of Field OB Stars in the Small Magellanic Cloud\footnotemark[1]}\footnotetext[1]{This paper includes data gathered with the 6.5 meter Magellan Telescopes located at Las Campanas Observatory, Chile.}


\author{J. B. Lamb, M. S. Oey, A. S. Graus, F. C. Adams, \& D. M. Segura-Cox}
\affil{Astronomy Department, University of Michigan,
    Ann Arbor, MI 48109}



\begin{abstract}
Some theories of star formation suggest massive stars may only form in clustered environments, which would create a deficit of massive stars in low density environments.  Observationally, Massey (2002) finds such a deficit in samples of the field population in the Small and Large Magellanic Clouds, with an IMF slope of $\Gamma_{\rm IMF} \sim 4$.  These IMF measurements represent some of the largest known deviations from the standard Salpeter IMF slope of $\Gamma_{\rm IMF} = 1.35$.  Here, we carry out a comprehensive investigation of the mass function above $20M_\odot$ for the entire field population of the Small Magellanic Cloud, based on data from the Runaways and Isolated O Type Star Spectroscopic Survey of the SMC (RIOTS4).  This is a spatially complete census of the entire field OB star population of the SMC obtained with the IMACS multi-object spectrograph and MIKE echelle spectrograph on the Magellan telescopes.  Based on Monte-Carlo simulations of the evolved present-day mass function, we find the slope of the field IMF above $20M_\odot$ is $\Gamma_{\rm IMF}$=2.3$\pm 0.4$. We extend our IMF measurement to lower masses using BV photometry from the OGLE II survey.  We use a statistical approach to generate a probability distribution for the mass of each star from the OGLE photometry, and we again find $\Gamma_{\rm IMF}$=2.3$\pm 0.6$ for stellar masses from 7$M_\odot$ to 20$M_\odot$.  The discovery and removal of ten runaways in our RIOTS4 sample steepens the field IMF slope to $\Gamma_{\rm IMF}$=2.8$\pm 0.5$.  We discuss the possible effects of binarity and star-formation history on our results, and conclude that the steep field massive star IMF is most likely a real effect.
\end{abstract}





\section{Introduction}

In many ways, the mass of a star is its most important attribute.  A star's mass constrains not just its observable properties and future evolution, but also provides an observational link to the local conditions under which it formed.    Similarly, the initial mass function (IMF) is a fundamental parameter of a stellar population and provides a direct probe of the star formation process.  

The form of the high mass tail of the stellar initial mass function (IMF) was first described by Salpeter (1955), who found it could be well-described by a simple power law of the form 
$n(m)\ dm \propto m^{-\gamma} dm$ where $n$ is the number of stars, $m$ is stellar mass, and $\gamma$ = 2.35.  This original measurement has proven to be robust for stars $\gtrsim 1M_\odot$, with stellar populations from small OB associations to young, massive clusters all exhibiting this canonical Salpeter IMF (Kroupa 2001, and references therein).  These widespread similarities in the top end of the IMF from such disparate populations may imply that the IMF is a universal property of star formation, regardless of environment (e.g., Elmegreen 2000).  However, the universality of the IMF is still an important open question (see Bastian et al. 2010).  

Alongside the power law slope of the IMF, the stellar upper mass limit, $\mup$, is also a critical component of a stellar population.  Studies of well-populated clusters indicate $\mup \sim 150M_\odot$ (e.g., Oey \& Clarke 2005), although masses up to twice as large are reported (e.g., Crowther et al. 2010).  However, $\mup$ for small-scale, isolated star formation is poorly constrained.

The field star population is an ideal target for an investigation into the consistency of the Salpeter IMF slope and $\mup$ for distributed, sparse star formation. Furthermore, the ability of massive stars to form in isolation is a distinguishing test between two popular theories of massive star formation, monolithic collapse (e.g., Shu et al. 1987) and competitive accretion (e.g., Zinnecker 1982).
In the monolithic collapse model, molecular clouds fragment unevenly into clumps that will each form a single star, which will accrete material solely from its own fragment of the cloud.  In this model, the mass of the fragment determines the available mass to form the star, so massive stars will form from massive fragments.  In contrast, the competitive accretion model predicts that molecular cloud fragments are not limited to their own gas mass, but rather accrete from a shared reservoir of gas in the molecular cloud.  In this scenario, high mass stars preferentially form in the dense centers of molecular clouds, where more gas available for accretion exists.  From these two models, only the former is compatible with field massive star formation, although the specific mechanism that would allow a small molecular cloud to avoid fragmentation altogether is unclear.  3-D hydrodynamic simulations by Krumholz et al. (2009) reveal a scenario where a high-mass star can form in isolation or alongside a few low mass stars.  In contrast, simulations of competitive accretion indicate a specific correlation between cluster mass ($\Mcl$) and the most massive star a cluster will form ($\mmax$), given by $\Mcl \propto \mmax^{1.5}$ (Bonnell et al. 2004).  Thus, according to the competitive accretion scenario, massive stars are incapable of forming in isolation.

The concept of a $\mmax$-$\Mcl$ relation is also advocated by Weidner \& Kroupa (2006), who use analytical models and an aggregation of Galactic cluster data to support a deterministic $\mmax$-$\Mcl$ relationship, similar to that of competitive accretion.  Using an expanded Galactic cluster dataset, Weidner et al. (2010) argue that it is statistically improbable that these clusters were randomly populated from the same universal stellar IMF and that a clear relationship between $\mmax$ and $\Mcl$ exists.  One of the primary consequences of a deterministic $\mmax$-$\Mcl$ relationship is that the integrated galaxial initial mass function (IGIMF) would necessarily be steeper than the canonical Salpeter IMF (Weidner \& Kroupa 2005), since the most massive stars are restricted to forming only in the most massive of clusters.  Similarly, the formation of massive stars in isolation appears contradictory to this model.  

In situ formation is not the only explanation for isolated massive stars.  There are a number of methods by which stars formed in clusters may appear in isolation.  Runaway stars, in particular, are a well-established component of the field massive star population.  These runaways are stars formed in the dense cores of clusters, which are ejected from their birth cluster either dynamically (Poveda et al. 1967) or by receiving a kick from a supernova explosion (Blaauw 1961).  Estimates vary greatly on their fractional contribution to the field population, with observed values between 10\% (Blaauw, 1961) and 50\% (de Wit et al. 2005).  While these works identify runaways using their high peculiar space velocities ($>$ 30 km/s), the existence of slow runaways (Banerjee et al. 2012) or two-step ejections that reduce space velocities (Pflamm-Altenburg \& Kroupa 2010) may result in such runaway fractions being underestimated.  Gvaramadze et al. (2012) suggests that the runaway fraction may be as high as 100\%, considering the observational difficulties of identifying low velocity runaways.  Another potential origin of field massive stars are clusters that quickly expel their gas, which may cause rapid dissociation of the cluster, ``infant mortality" (e.g., de Grijs \& Goodwin 2008) or a large fraction of it, ``infant weight loss" (e.g., Bastian \& Goodwin, 2006).    Finally, in some cases, sparse clusters may simply exist undetected around field massive stars (e.g., Lamb et al. 2010).  As a whole, the field population may be a combination of some or all of these separate stellar populations.  

The heterogeneous nature of the field population significantly complicates the determination of the mass function.  Each component of the field may have a different impact on the stellar mass function.  The frequency of runaways, for example, correlates with spectral type (e.g., Blaauw 1961).  Stone (1991) found runaway fractions of 30-40\% and 5-10\% for O and early B stars, respectively.  Thus, the runaway population would tend to flatten the observed present day mass function (PDMF) of the field.  In contrast, competitive accretion theory and the $\mmax$-$\Mcl$ relation suggest that sparse massive star formation or in situ field formation may not sample the top end of the IMF, which would result in steepening the field IMF in a manner similar to the proposed steepening of the IGIMF.  However, the mass function of field stars formed in situ is an unknown quantity and may be a product of a different mode of star formation, distinct from clustered star formation.

Studies of field massive star populations have yet to converge on a value for its stellar mass function.  An analysis by van den Bergh (2004) directly compares the spectral types of clustered stars versus field stars from a survey of Galactic O stars (Ma\'iz-Apell\'aniz et al. 2004).  He finds that field OB stars are skewed towards later spectral types than their clustered counterparts, which suggests that the field population is either less massive or older than the cluster population.  Massey et al. (1995) and Massey (2002) use a combination of spectroscopy and photometry to measure the high mass stellar IMF of clusters, associations and a few sample field regions in the Small and Large Magellanic Clouds.  While the clusters and associations exhibit a standard Salpeter IMF slope, $\Gamma = \gamma -1 =$ 1.35, the field IMF is significantly steeper, with a slope of $\Gamma \sim$ 4 above their completeness limit of $25M_\odot$.  However, Selman et al. (2011) show that the field population around the 30 Dor region of the Large Magellanic Cloud is consistent with a Salpeter IMF.  Intermediate values have also been found, with Zaritsky et al. (1998) find $\Gamma$ = 1.8 for the LMC field from $7M_\odot$-$35M_\odot$ using photometry from the Ultraviolet Imaging Telescope.  Similarly, {\'U}beda et al. (2007) derive an IMF slope of $\Gamma$ = 1.8 for NGC 4214 using  {\it HST} WFPC2 and STIS photometry.  Resolving these discrepancies requires a robust determination of the field massive star IMF using a reliable estimator for mass, such as spectroscopy.

The field studies discussed above fall into two general categories: a small-scale survey using reliable, spectroscopic mass estimates, or a large-scale photometric survey that yields less reliable masses.  Spectroscopic studies in particular have significant completeness issues and are limited to the nearby Galactic field or sub-samples of nearby galaxies.  These limitations lead to an incomplete picture of the field massive star population.  In this work, we present the first spatially complete, spectroscopic survey of an entire galaxy's field massive star population.  We target the Small Magellanic Cloud (SMC) in our Runaways and Isolated O Type Star Spectroscopic Survey of the SMC (RIOTS4).  With RIOTS4 spectra, we obtain accurate masses for the entire population of field massive stars across the full spatial extent of the SMC.  This unprecedented data set will yield a definitive mass function for field massive stars in the SMC, including the slope and upper mass limit.  We discuss the importance of these results in the context of massive star formation models and highlight differences between the field and clustered populations of massive stars.

\section{RIOTS4}

Here, we present an abbreviated description of the RIOTS4 survey.  A complete description can be found in Lamb et al. (in prep).
RIOTS4 targets the complete sample of 374 SMC field OB stars, as identified in Oey et al. (2004).  Selection of these targets is a two step process.  First, Oey et al. (2004) identified massive stars in the SMC using two photometric selection criteria, $B \leq 15.21$ and $Q_{UBR} \leq -0.84$, where $Q_{UBR}$ is the reddening free parameter given by
\begin{eqnarray}
Q_{UBR} &=& (m_U - m_R) - \frac {A_U - A_R} {A_B - A_R}(m_B - m_R) \nonumber \\
&=&(m_U - m_R) - 1.396(m_B - m_R)\ .
\end{eqnarray}

Photometry from Massey (2002) is used for this selection.  Second, the field stars are selected by running a friends-of-friends algorithm (Battinelli 1991) on the massive star sample.  This algorithm identifies stellar clustering by setting a physical clustering length such that the number of clusters is maximized.  Thus, our RIOTS4 targets are those OB stars that are at least one clustering length (28 pc) removed from any other OB stars.

The primary instrument for RIOTS4 is the Inamori-Magellan Areal Camera and Spectrograph (IMACS) on the Magellan Baade telescope at the Las Campanas Observatory.  One of the primary benefits of IMACS is the capability of multi-object observations using slitmasks.  We observe a total of 328 objects in 49 slitmasks in the f/4 observing mode.  The multi-object observing setup is designed to maximize spectral resolution with a 1200 lines/mm grating and either a 0.7" or 1" slit width, resulting in spectral resolutions of R $\sim$ 3700 and R $\sim$ 2600, respectively.  
Wavelength coverage varies between spectra, but every spectrum includes coverage from 4000-4700 $\AA$.  The exposure time for each multi-object observation is one hour, split into three 20 minute exposures. We conducted all multi-object observations between September 2006 and December 2010.  

We were unable to fit all stars onto multi-slit masks, due to the density of targets being too high to fit all the objects onto the mask or too low to warrant the use of multi-object slitmasks.  In these cases, we instead observe targets using IMACS long slit observations, or with the Magellan Inamori Kyocera Echelle (MIKE) single object spectrograph on the Magellan Clay telescope.  With IMACS, we observe 27 stars with a 300 l/mm grism in f/2 mode with 0.5" - 0.7" slit widths, which yield spectral resolutions of R $\sim$ 1000 - 1300.  As before, these observations are three 20 minute exposures.   With MIKE, we observe 48 stars with a 1" slit width, which yields R $\sim$ 28000.  For these observations, exposure times range from 15 - 30 minutes depending on the brightness of the target and are designed to achieve a S/N $> 30$.  We performed these observations between November 2010 and October 2011.

To reduce the multi-object spectra, we use the Carnegie Observatories System for MultiObject Spectroscopy (COSMOS) data reduction package\footnote{COSMOS was written by A. Oemler, K. Clardy, D. Kelson, G. Walth, and E. Villanueva.  See http://code.obs.carnegiescience.edu/cosmos.}.  COSMOS is specifically designed to reduce and extract IMACS spectra spread across eight CCD chips.  We follow the standard COSMOS cookbook to perform bias-subtraction, flat-fielding, wavelength calibration and extraction of 2-D spectra.  With the 2-D spectra output from COSMOS, we use the apextract package in IRAF\footnote{IRAF is distributed by the National Optical Astronomy Observatory, which is operated by the Association of Universities for Research in Astronomy (AURA), Inc., under cooperative agreement with the National Science Foundation (NSF).} to find, trace, and extract stellar apertures, which yields 1-D spectra.  We reduce MIKE and IMACS long slit spectra with standard IRAF procedures and use the apextract package to generate 1-D spectra.  With 1-D spectra in hand, we rectify them using the continuum procedure and eliminate remaining cosmic rays or bad pixel values with the lineclean procedure, which are both part of the onedspec package in IRAF.

\section{Spectral Catalog}
\label{catalogsec}

\begin{figure*}
	\begin{center}
	\includegraphics[scale=.8,angle=90]{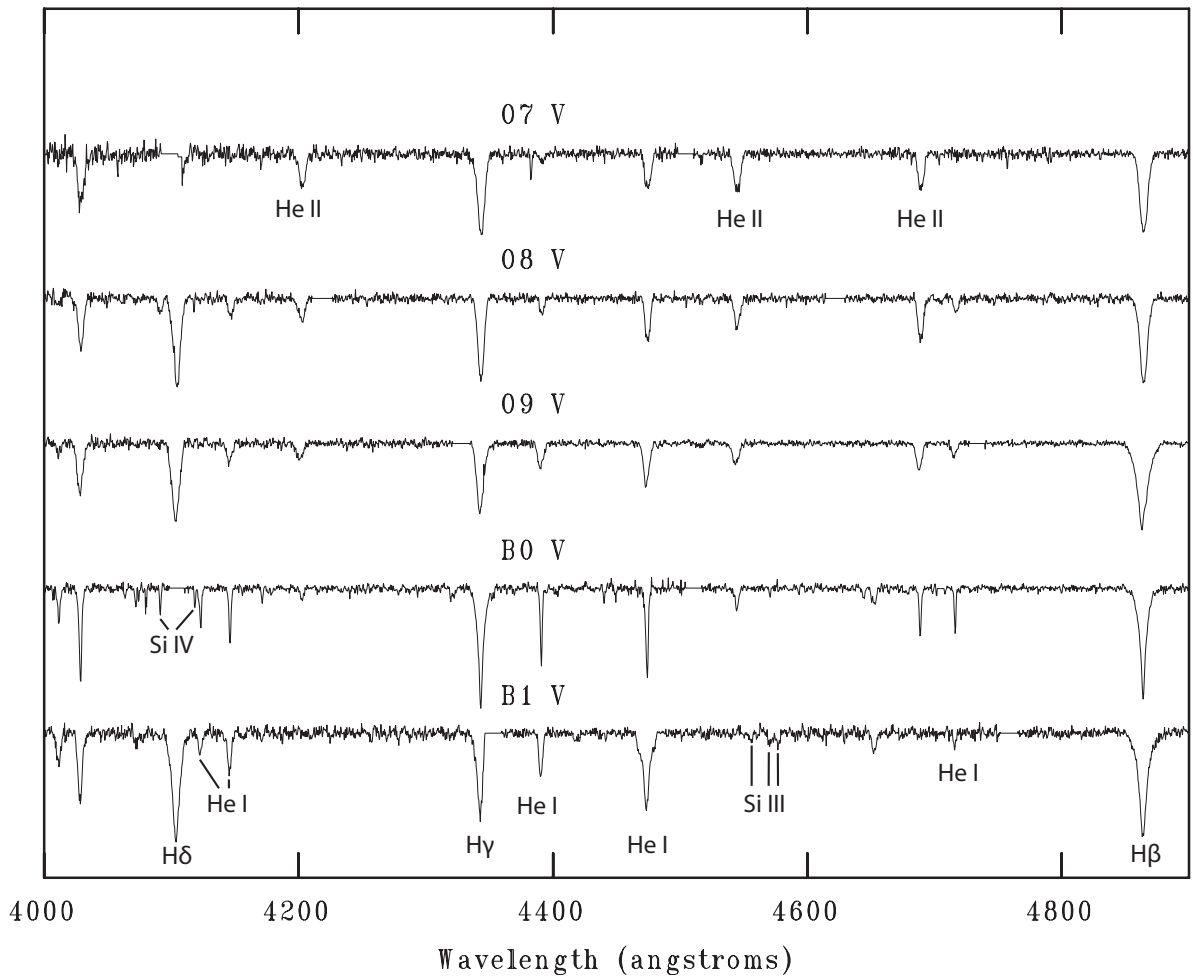}
	\caption{A sequence of spectral types from 07 V to B1 V from the RIOTS4 survey.  We label the major spectral features in the range from $4000 - 4900 \AA$.  The ratio of He {\footnotesize II} $\lambda$4542 to He {\footnotesize I} $\lambda$4471 is a primary spectral type diagnostic for O stars.  Notice how this ratio decreases towards later spectral types, until the disappearance of He {\footnotesize II} at B1 V.  }
	\label{sequence}
	\end{center}
\end{figure*}

The key observational result of the RIOTS4 survey is the distribution of stellar spectral types.  For field massive stars, the number of stars and completeness of the sample are both unprecedented in spectroscopic studies.  We obtain spectral types for stars in the RIOTS4 survey using a qualitative comparison with the spectral atlas of Walborn \& Fitzpatrick (1990), with additional reference to Lennon (1997), Walborn et al. (2000), Walborn et al. (1995)  and Walborn (2009).  J. B. L., M. S. O., and A. S. G. individually assign spectral types for each star.  Any discrepancies are re-examined until a consensus is reached.  We polish our catalog further by plotting spectra sequentially according to spectral type in an iterative process to more clearly define the boundaries between spectral types.  The majority of our spectral types accurate to within half a type; thus, a star cataloged as O9 may range from O8.5 - O9.5.  For stars with more uncertainty, we list a range of spectral types in our catalog. 

In Figure \ref{sequence}, we plot a sequence of RIOTS4 spectra, which cover spectral types from O7 V to B1 V.  The primary diagnostic lines used for spectral typing are the ratio of He {\footnotesize II} $\lambda$4542 to He {\footnotesize I} $\lambda$4471 for O stars (see Figure \ref{sequence}) and the ratio of Si {\footnotesize IV} $\lambda$4088 to Si {\footnotesize III} $\lambda$4555 for B stars.  Luminosity classes are determined using an iterative approach, where spectral diagnostics are the primary criterion and photometric magnitudes are a secondary criterion.  Primary spectral diagnostics for luminosity class include emission of N {\footnotesize II} $\lambda\lambda$4634-4640-4042 and absorption/emission of He {\footnotesize II} $\lambda$4686 for stars earlier than O8, the ratio of Si {\footnotesize IV} $\lambda$4088 to He {\footnotesize I} $\lambda$4026 for late O stars, and the ratio of Si {\footnotesize III} $\lambda$4555 to He {\footnotesize I} $\lambda$4471 for B stars.  Due to the lower metallicity of the SMC, many stars in our sample exhibit weak metal lines in comparison to the Walborn \& Fitzpatrick (1990) catalog.  Thus, for evolved stars, we also rely heavily on the classification criterion established by Lennon (1997) for SMC supergiants.  Finally, photometric magnitudes (Massey 2002) do provide an additional check on the luminosity class.  However, multiplicity cannot be ruled out and should be expected at a significant level.  As part of the RIOTS4 survey, we investigate the binary fraction of SMC field massive stars by taking $\sim$10 epochs of observations for 30 stars our sample.  Preliminary results from these observations indicate a binary fraction $\gtrsim 50\%$ (Lamb et al. in prep), which is similar to that found in open clusters (e.g Sana et al. 2008, 2009, 2011).

\begin{figure*}
	\begin{center}
	\includegraphics[scale=.8,angle=90]{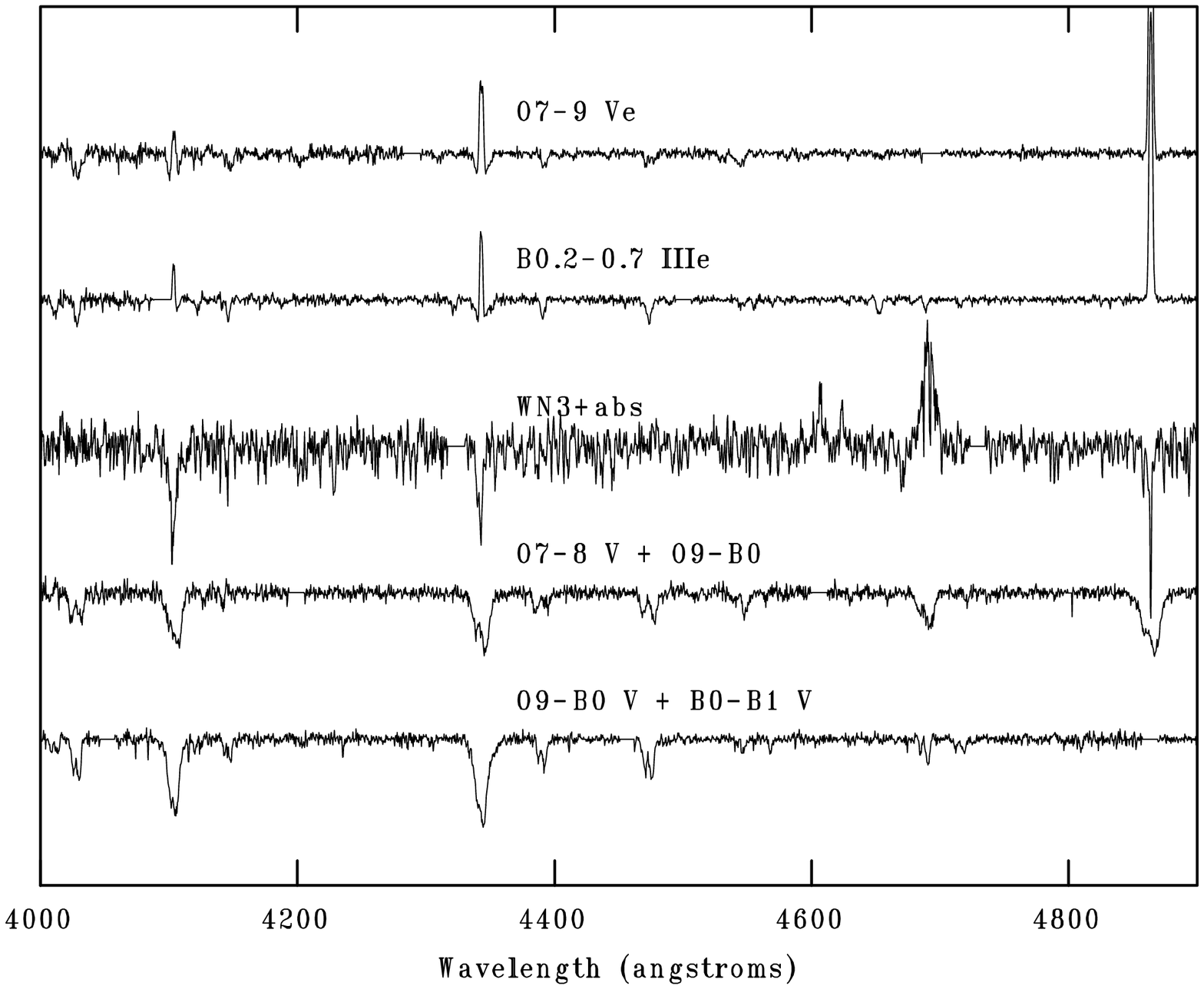}
	\caption{A collection of binary and emission line stars from the RIOTS4 survey.  From top to bottom, these are examples of an Oe star, a Be star, a Wolf-Rayet star, an O+O binary system, and an O+B binary system.  The spectral type ranges for the Oe and Be star include a direct measurement based on line ratios and an estimate of the equivalent photospheric spectral type the star would have if infilling of He {\footnotesize I} did not occur.}
	\label{OeBeBinary}
	\end{center}
\end{figure*}

Two populations add considerable difficulties to the RIOTS4 spectral catalog: binaries and emission line stars (see Figure \ref{OeBeBinary}).  We identify binaries either through double-peaked absorption lines, or the presence of two strong spectral lines that cannot originate from a single star, such as He {\footnotesize II} $\lambda$4542 and Si {\footnotesize III} $\lambda$4555.  To obtain spectral types for the binary population, we create a sequence of model binary star spectra, ordered by spectral type of the primary and secondary objects.  To avoid issues with metallicity, we create these model binaries directly from the RIOTS4 data.  First, we median combine RIOTS4 spectra with identical spectral types to obtain a template spectrum for that type.  Then, we combine two template spectra, appropriately accounting for luminosity, to generate a model binary star.  In this manner, we find that we can identify the primary object to within one spectral type in most cases.  However, the secondary star is rarely well constrained, especially in the case of single-lined binaries.  The spectral types of B star companions to a primary O stars, which represent the majority of our binary systems, are particularly difficult to determine.  These difficulties are due in large part to the relatively weak metal lines that distinguish B star spectral types, and the luminosity difference between O and B stars.  We also investigate the consequences of a large population of undetected binaries in \S \ref{binarysec}.

The second troublesome population are emission line (Wolf-Rayet and Oe/Be) stars.  Our survey includes a pair of well studied Wolf-Rayet (WR) stars, for which we adopt their physical parameters from Massey (2002).  However, for Oe/Be stars, their emission lines arise from hot circumstellar material around the star.  In the case of classical Oe/Be stars, this material exists in a `decretion disk' caused by the rapid rotation of the star (e.g., Porter \& Rivinius 2003).  This population is expected, since classical Oe/Be stars are more common in the SMC than in the Galaxy or the LMC (e.g., Bonanos, 2010).  Classical Oe/Be stars represent a significant fraction of the RIOTS4 sample, with $\sim 50$\% of B stars and $\sim 20$\% of O stars exhibiting emission in one or more Balmer lines.  However, these fractions are artificially high due to the selection criteria for RIOTS4.  Classical Oe/Be stars exhibit strong H-$\alpha$ emission, which brightens the $R$ magnitudes of these stars.  In turn, this effect lowers the value of $Q_{UBR}$ in these objects, which causes them to be preferentially included in the RIOTS4 survey due to our $Q_{UBR} \leq -0.84$ selection criterion.  
Thus, the completeness of Be stars extends to slightly later spectral types than for normal B stars in the RIOTS4 survey.  For measuring stellar masses, the primary issue with Oe/Be star spectra is the presence of weak or filled-in absorption lines.  Spectral types are significantly impacted when the important diagnostic lines listed above appear filled-in or non-existent.  In the particular case of Oe stars, infilling does not affect the He {\footnotesize II} lines but may impact He {\footnotesize I} lines, which may bias these stars to earlier spectral types (Negueruela et al. 2004).  To deal with this issue, we adopt a large spectral type range for Oe stars that show evidence of infilling in He {\footnotesize I} lines.  The earliest type in this range is obtained from the ratio of He {\footnotesize II} to He {\footnotesize I} line strengths, as in non-emission line stars.  The later type in this range is estimated by the relative strengths of the different He {\footnotesize II} lines and indicates the spectral type the star would have without infilling.  This equivalent photospheric spectral type allows us to obtain a better estimate of $T_{\rm eff}$.  Oe/Be stars with smaller ranges indicate objects where the important diagnostic lines appear with smooth, Gaussian profiles.  Ultimately, we adopt the median spectral type from these ranges in our derivation of their stellar parameters, but adopting either the early or later types in these ranges does not significantly impact our results (see \S \ref{riotsimfsec}).

There is some overlap of our survey with other spectroscopic studies of the SMC; however, our typical S/N $\sim$ 75 and resolution R $\sim 3000$ compare favorably to these studies.  A number of our targets were observed by Massey et al. (1995) with similar S/N $\sim$ 75 but lower resolution (R $\sim$ 1500).  We find good agreement with our spectral types.  When discrepancies do arise, they are always within half a spectral type or two luminosity classes.  Another study that significant overlaps with RIOTS4 is the 2dF survey of the SMC (Evans et al. 2004; Evans \& Howarth 2008).  Their spectroscopic data is slightly lower quality than RIOTS4, with average S/N $\sim$ 45 and R $\sim$ 1600.  Our agreement with 2dF is not as good as with Massey et al. (1995), with the majority of discrepancies arising with luminosity class.  In general, we agree with 2dF spectra to within one spectral type and two luminosity classes.  Evans et al. (2004) classify a large fraction of the overlapping sample of stars as giants, many of which we classify as dwarfs.  However, due to the poor quality of their spectra, they rely on a combination of the equivalent width of H-$\gamma$ and stellar magnitude for their luminosity classifications.  Thus, their ad hoc methodology may explain this apparent discrepancy in luminosity classifications.

\section{HR Diagram and Stellar Masses}
\label{HRDsection}

We use our spectral types and photometry from Massey (2002) to derive the physical properties of 284 individual stars in RIOTS4.  We exclude stars with no diagnostic lines (mainly Oe/Be) from this analysis.  In Table \ref{catalogtable}, we list the spectral types and physical properties for all stars in the RIOTS4 survey that contribute to our measurement of the IMF ($> 20M_\odot$).  We list the ID number, $B$ magnitude, and $V$ magnitude from Massey (2002) in columns 1, 2, and 3, respectively.  Photometric errors are typically $\sim 0.03$ mag in $B$ and $\sim 0.02$ mag in $V$ (Massey 2002).  Column 4 lists $Q_{UBR}$ calculated from the Massey (2002) photometry.  We list bolometric magnitude, $M_{\rm Bol}$, in column 5, which is calculated using the extinction, $A_V$, and stellar effective temperature, $T_{\rm eff}$ (Section \ref{HRDsection}, given in columns 6 and 7, respectively).  Finally, we list the estimated mass and observed spectral type from RIOTS4 in columns 8 and 9, respectively.  Stars with uncertain spectral types are given a range, with their adopted $T_{\rm eff}$ coming from the median spectral type within that range.  To obtain stellar effective temperature, $T_{\rm eff}$, we use two different calibrations, one for O stars and one for B stars.  Due to the lower metallicity in the SMC (e.g., Hunter et al. 2007), the $T_{\rm eff}$ of O stars is systematically higher than in the Galaxy for stars of the same spectral type (Massey et al. 2005).  For O stars, we therefore convert spectral types to $T_{\rm eff}$ using the calibration in Massey et al. (2005) for the SMC.  For B stars, we use conversions to $T_{\rm eff}$ from Crowther et al. (1997).  These calibrations overlap smoothly at a spectral type of B0.  Although Crowther et al. (1997) use a sample of Galactic stars for their calibration, Massey et al. (2005) demonstrate that $T_{\rm eff}$ for SMC and Galactic stars is equal for stars B0 and later.  We calculate bolometric magnitudes using $M_{\rm Bol}$ = $V$ - $DM$ - $A_V$ + $BC$, where $DM$ is the distance modulus, $A_V$ is the extinction, and $BC$ is the bolometric correction.  We adopt $DM = 18.9$ (Harries et al. 2003) and  $BC = 27.99 - 6.9\log T_{\rm eff}$ (Massey et al. 2005).  $A_V$ is found using the SMC extinction maps from the Magellanic Clouds Photometric Survey (MCPS; Zaritsky et al. 2002).  

\begin{figure*}
	\begin{center}
	\includegraphics[scale=,angle=0]{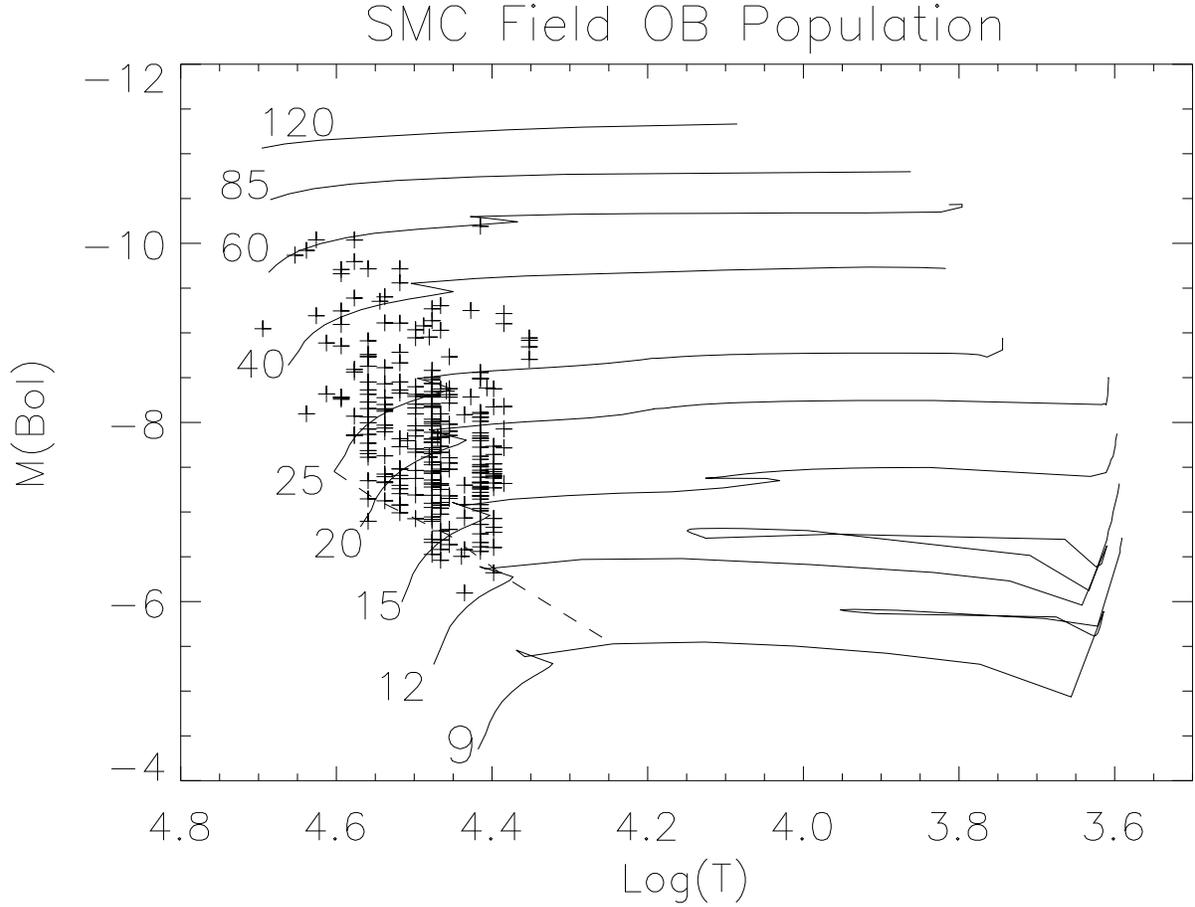}
	\caption{An H-R diagram of our field stars with physical parameters $T_{\rm eff}$ and $M_{\rm Bol}$ derived from spectral types and Massey (2002) photometry.  The evolutionary tracks plotted here are from Charbonnel et al. (1993) at SMC metallicity and are labelled corresponding to the stellar mass of the evolutionary track in $M_\odot$.  The dashed line indicates the completeness limit of RIOTS4.}
	\label{photcmd}
	\end{center}
\end{figure*}

\begin{deluxetable*}{ccccccccc}
 \tabletypesize{\small}
  \tablewidth{0pc}
  \tablecaption{RIOTS4 Spectral Catalog\tablenotemark{a}} 
\tablehead{\colhead{ID} & \colhead{B} & \colhead{V} & \colhead{M$_{\rm bol}$} & \colhead{Q} & \colhead{A$_{\rm V}$} & \colhead{T$_{\rm eff}$ (K)} & \colhead{Mass ($M_\odot$)} &\colhead{Sp Type} }
\startdata

 1600  & 14.42 & 14.60 & -11.15 &  -0.87 &   0.32 & 33000 & 23.6 $^{+1.7} _{-1.5}$ & O9 V  \\  \vspace{2pt}
 3459  & 13.32 & 13.46 & -12.30 &  -0.93 &   0.48 & 33000 & 35.9 $^{+3.3} _{-5.0}$ & O9 III  \\  \vspace{2pt}
 4919  & 13.66 & 13.85 & -11.34 &  -0.95 &   0.33 & 29250 & 26.2 $^{+4.1} _{-4.7}$ & B0.2 III  \\  \vspace{2pt}
 5313  & 14.89 & 15.11 & -10.78 &  -0.87 &   0.23 & 34500 & 20.9 $^{+1.8} _{-1.5}$ & O8.5 V  \\  \vspace{2pt}
 7437  & 12.93 & 13.12 & -14.32 &  -0.94 &   0.33 & 42250 & 75.0 $^{+5.4} _{-4.7}$ & O6 III((f))  \\  \vspace{2pt}
 7782  & 14.30 & 14.46 & -12.30 &  -0.91 &   0.40 & 37750 & 33.2 $^{+2.4} _{-2.2}$ & O7.5 V  \\  \vspace{2pt}
 9732  & 14.63 & 14.81 & -11.63 &  -0.88 &   0.31 & 39250 & 27.1 $^{+2.7} _{-2.2}$ & O7 Vz  \\  \vspace{2pt}
11045  & 14.80 & 15.01 & -11.37 &  -0.87 &   0.21 & 34500 & 24.9 $^{+2.6} _{-1.8}$ & O8.5 V  \\  \vspace{2pt}
11623  & 14.12 & 14.13 & -11.79 &  -0.86 &   0.83 & 33000 & 28.4 $^{+3.4} _{-1.2}$ & O9 V  \\  \vspace{2pt}
11677  & 14.47 & 14.46 & -10.82 &  -1.01 &   1.04 & 33000 & 21.0 $^{+1.6} _{-1.4}$ & O9 V  \\  \vspace{2pt}
\enddata
\tablenotetext{a}{This table is published in its entirety in the electronic edition of the $Astrophysical$ $Journal$.  A portion is shown here for guidance regarding its form and content.}
\label{catalogtable}
\end{deluxetable*}

With $T_{\rm eff}$ and $M_{\rm Bol}$ computed, we construct a Hertzprung-Russell diagram of the SMC field massive star population (Figure \ref{photcmd}).  Stars from the RIOTS4 survey are plotted as plus signs, while the lines represent Geneva stellar evolutionary tracks at a metallicity of Z=0.004, consistent with the SMC (Charbonnel et al. 1993).  These tracks are labelled by stellar mass in $M_\odot$.  We plot the completeness limit of RIOTS4 as a dashed line.  Thus, RIOTS4 is complete to 20 $M_\odot$ along most of the main sequence and 25 $M_\odot$ along the ZAMS.  Although relatively few stars are observed below the completeness limit, this is due to the RIOTS4 selection criteria, rather than an indication of the completeness of Massey (2002) photometry.  Due to these completeness issues, we will limit our measurement of the field IMF in Section \ref{riotsimfsec} to stars $> 20$ $M_\odot$.  One striking feature of this HR diagram is the shift of the observed main sequence from the main sequence of the Geneva models.  There is a distinct lack of stars observed along the modeled ZAMS and likewise, a significant population of objects extending past the main sequence turn-off of the models.  A similar distribution is seen in a sample of SMC field stars from Massey (2002) (his Figure 10) and in the SMC cluster NGC 346 from Massey et al. (1995) (their figure 8).  Thus, it is unclear if this offset is a real property of the SMC, or due to some systematic issue with either the Geneva models or the calibration and calculation of $T_{\rm eff}$ and $M_{\rm Bol}$.  One possible explanation is that the Geneva models plotted here are older, non-rotating models.  We opt to use the non-rotating models so we can directly compare the stellar IMF for the mass ranges covered by both the RIOTS4 survey ($> 20$ $M_\odot$) and OGLE photometry ($7 - 20$ $M_\odot$).  We note that the models with rotation at SMC metallicity do shift the main sequence to cooler $T_{\rm eff}$, but the magnitude of the effect is only $\sim$ 0.04 dex for stars rotating at 300 km s$^{-1}$ (Maeder \& Meynet 2001).  Thus, rotation alone cannot explain the discrepancy shown in Figure \ref{photcmd}.  Therefore, it appears that SMC field massive stars are systematically cooler or more evolved than expected from the Geneva stellar evolutionary models.  

Despite the above issues with the Geneva models, the nearly horizontal evolution of stars off the main sequence in $M_{\rm Bol}$ vs. $T_{\rm eff}$ helps to mitigate their impact on stellar mass estimates.
Thus, with these Geneva models and the derived stellar parameters of $T_{\rm eff}$ and $M_{\rm Bol}$, we proceed to estimate the stellar mass of each individual star.  To accomplish this, we linearly interpolate between the Geneva model isochrones (Charbonnel et al. 1993) to match the stellar parameters.  The primary source of error in our mass estimates is due to uncertainty in our spectral types.  Our typical uncertainty of half a type corresponds to $\sim$1500 K or 1-5 $M_\odot$, depending on mass of the star.  Stars with higher masses will typically have larger errors due to the spacing of tracks in the $M_{\rm Bol}$ vs. $T_{\rm eff}$ parameter space.  Another potential source of error is the discrepancy between the models and observations discussed above.  However, since this appears to be a systematic effect, its impact on the shape of the IMF should be small.  

\section{The Field Massive Star IMF}
\label{imfsec}

\subsection{The Field IMF above $20M_\odot$}
\label{riotsimfsec}

\begin{figure*}
	\begin{center}
	\includegraphics[scale=.65,angle=0]{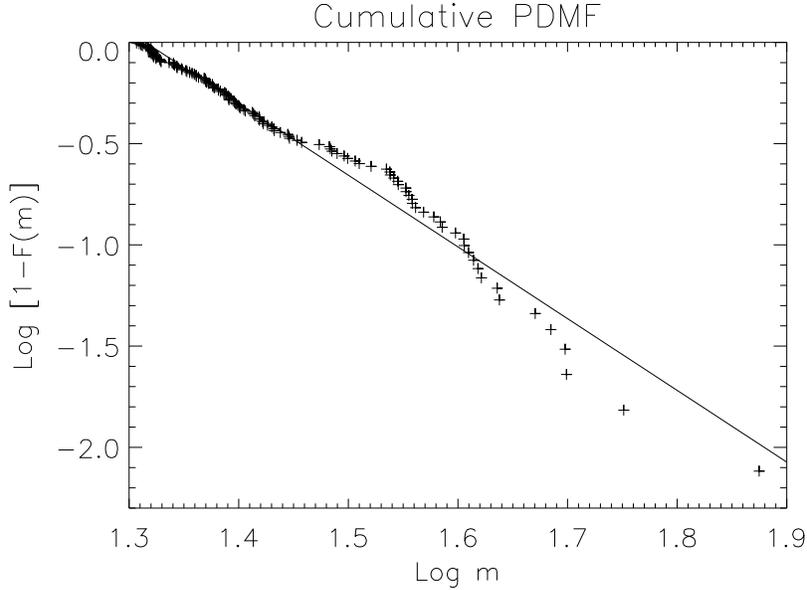}
	\caption{The PDMF of the SMC field star population, plotted as $\log[1-F(m)]$ versus $\log m$, where $F(m)$ is the empirical CDF.  The plotted line shows the linear least squares fit to the data, with a slope of $\Gamma_{\rm PDMF}$=3.5.}
	\label{koen}
	\end{center}
\end{figure*}

We proceed with a derivation of the stellar mass function of the field following the method of Koen (2006).  Since the field population is not coeval, we are actually measuring its present day mass function (PDMF), rather than its IMF.  Just as with the IMF, the PDMF can be described by a power law of the form
\begin{equation}
f(m) = \alpha m^{-(\Gamma_{\rm PDMF}+1)}\ ,
\end{equation}
where $\alpha$ is the normalization constant and $\Gamma_{\rm PDMF}$ is the logarithmic slope of the PDMF.  This PDMF can be described by a cumulative distribution function (cdf) of the form
\begin{eqnarray}
\label{cdfeq}
F(m) &=& \int_{\mlo}^m f(m')\, \mathrm{d}m' \nonumber \\
&=&  \frac {\alpha} {\Gamma_{\rm PDMF}} (\mlo^{-\Gamma_{\rm PDMF}} - m^{-\Gamma_{\rm PDMF}})\ ,
\end{eqnarray}
where $\mlo$ is the lower mass limit and $F(m)$ is the probability that a star's mass is between $\mlo$ and $m$.  
Normalization requires that the upper mass limit, $\mup$, follows $F(\mup) = 1$, which yields
\begin{equation}
\label{alphaeq}
\alpha = \frac {\Gamma_{\rm PDMF}} {\mlo^{-\Gamma_{\rm PDMF}}-\mup^{-\Gamma_{\rm PDMF}}}\ .
\end{equation}
From equations (\ref{cdfeq}) and (\ref{alphaeq}),
\begin{equation}
F(m) = \frac {1-(m/\mlo)^{-\Gamma_{\rm PDMF}}} {1-(\mup/\mlo)^{-\Gamma_{\rm PDMF}}}\ ,
\end{equation}
which can be written as
\begin{equation}
\label{slopeequation}
\log[1-\kappa F(m)] = -\Gamma_{\rm PDMF} \log m + \Gamma_{\rm PDMF} \log \mlo\ ,
\end{equation}
where $\kappa = [1-(\mup/\mlo)^{-\Gamma_{\rm PDMF}}]$.  In the case where $\mup \rightarrow \infty$ then $\kappa \rightarrow 1$ or if $\mup \gg \mlo$ then $\kappa \simeq 1$.
Following Koen (2006), the cdf $F(m)$ can be replaced by an empirical cdf given by
\begin{equation}
F[m(j)] = \frac {j} {(N+1)}\ ,
\end{equation}
where $N$ is the number of stars in the sample, $j$ is the rank of the star when the sample is ordered by increasing stellar mass.  Thus, 
$j$ goes from 1 to $N$, where 1 is the lowest mass star and $N$ is the highest mass star.  Using this empirical cdf, we generate a plot of $\log [1 - F(m)]$ vs. $\log m$ for the RIOTS4 sample (Figure \ref{koen}).  For this analysis, we adopt $\mlo$ = $20M_\odot$, which corresponds to the selection criteria for the RIOTS4 survey (see Figure \ref{photcmd}).  This limit yields a sample of 130 stars from which we will derive the PDMF.  It is clear from equation (\ref{slopeequation}) that when $\kappa \rightarrow 1$, $\Gamma_{\rm PDMF}$ is simply the slope of $\log [1 - F(m)]$ vs. $\log m$ from Figure \ref{koen}.  Thus, we use a linear least squares fit to obtain $\Gamma_{\rm PDMF}$=3.5 from Figure \ref{koen}.  

We also consider the case where $\kappa < 1$.  In this case, the form of the IMF assumes that of a truncated Pareto distribution given by
\begin{equation}
\log[1-F(m)] = \log \Big{[}1- \frac {1-(m/\mlo)^{-a}} {k}\Big{]} \ ,
\end{equation}
We perform two fits to the dataset assuming this form of the IMF, a nonlinear least squares and a maximum likelihood method, given by equations (8) and (10), respectively in Koen (2006).  From these fits, we find $\Gamma_{\rm PDMF}$= 2.8 using the non-linear least squares fitting and $\Gamma_{\rm PDMF}$= 3.1 using the maximum likelihood method.  

\begin{figure*}
	\begin{center}
	\includegraphics[scale=.65,angle=0]{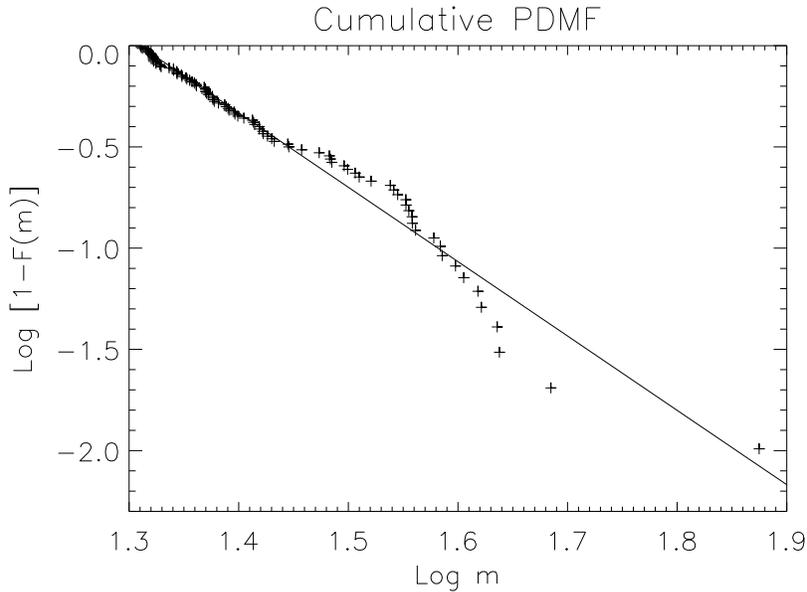}
	\caption{The PDMF of the SMC field star population with Oe/Be stars omitted, plotted as $\log[1-F(m)]$ versus $\log m$, where $F(m)$ is the empirical CDF.  The plotted line shows the linear least squares fit to the data, with a slope of $\Gamma_{\rm PDMF}$=3.7.}
	\label{koennoBe}
	\end{center}
\end{figure*}

Of the 130 stars contributing to this IMF measurement, Oe/Be stars account for 25\% (33) of these objects.  Since the spectral type and thus $T_{\rm eff}$ and mass of these stars is more uncertain, we test the impact these objects have on the PDMF slope.  As we discussed in \S \ref{catalogsec}, these stars have a larger possible range in their equivalent photospheric spectral type, which we assign individually for each star.  We find that adopting the earliest or latest extremes for these ranges changes the slope of the derived PDMF by 0.1 at most.  Adopting the earliest spectral types changes only the non-linear fit to $\Gamma_{\rm PDMF}$= 2.9, reducing the scatter in our PDMF measurements.  In contrast, adopting the latest spectral types increases the scatter, where $\Gamma_{\rm PDMF}$ = 3.6, 2.7, and 3.2 for linear and nonlinear least squares fits and maximum likelihood method, respectively.   Finally, if we exclude the Oe/Be stars entirely, the remaining 97 `normal' stars exhibit a slightly steeper slope of $\Gamma_{\rm PDMF}$= 3.7, 3.2, and 3.6 for linear and nonlinear least squares fits and maximum likelihood method, respectively (see Figure \ref{koennoBe}).  Thus, we find that the impact of Oe/Be stars on our derived mass function is small.

\begin{figure*}
	\begin{center}
	\includegraphics[scale=.65,angle=0]{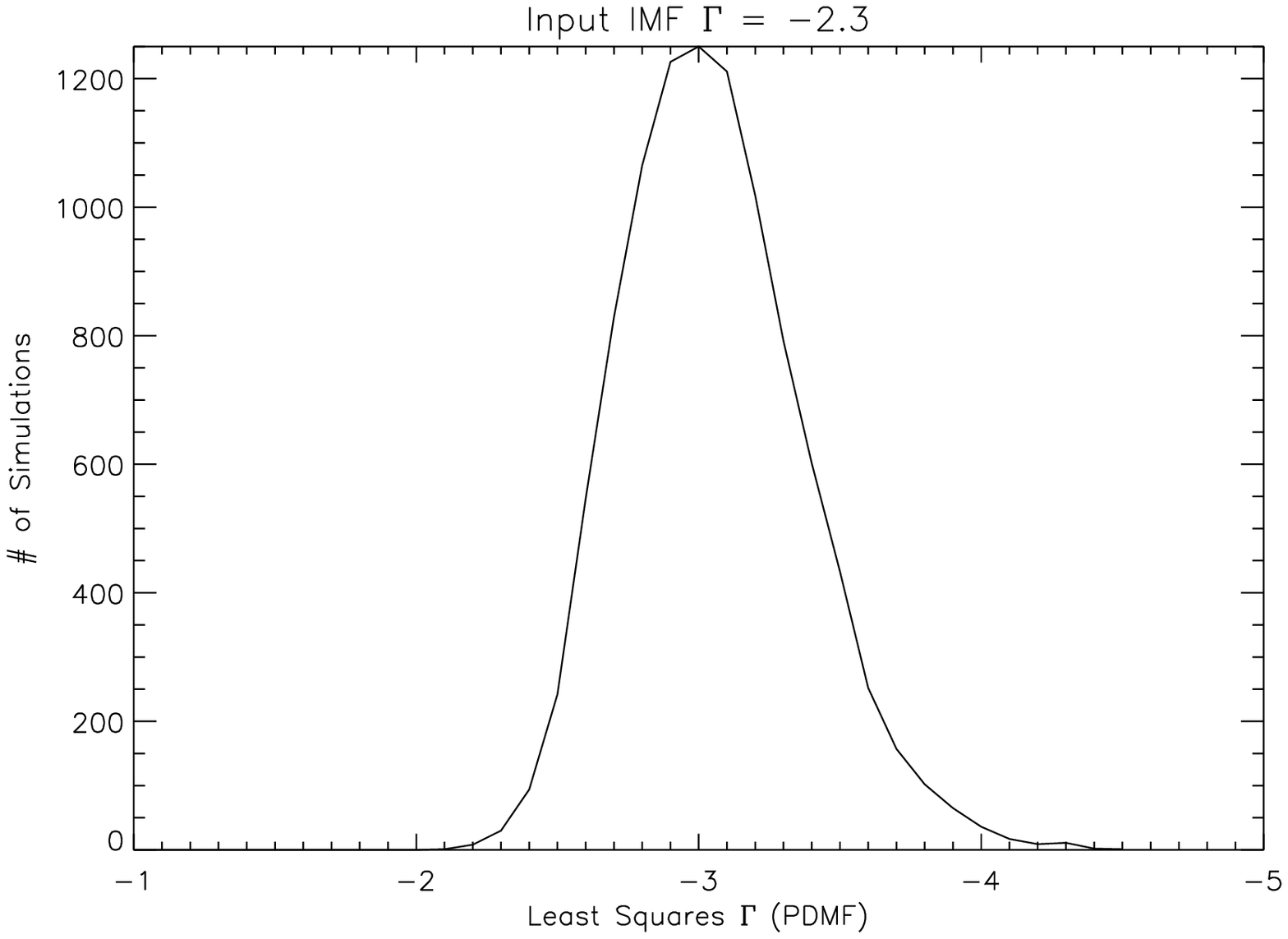}
	\includegraphics[scale=.65,angle=0]{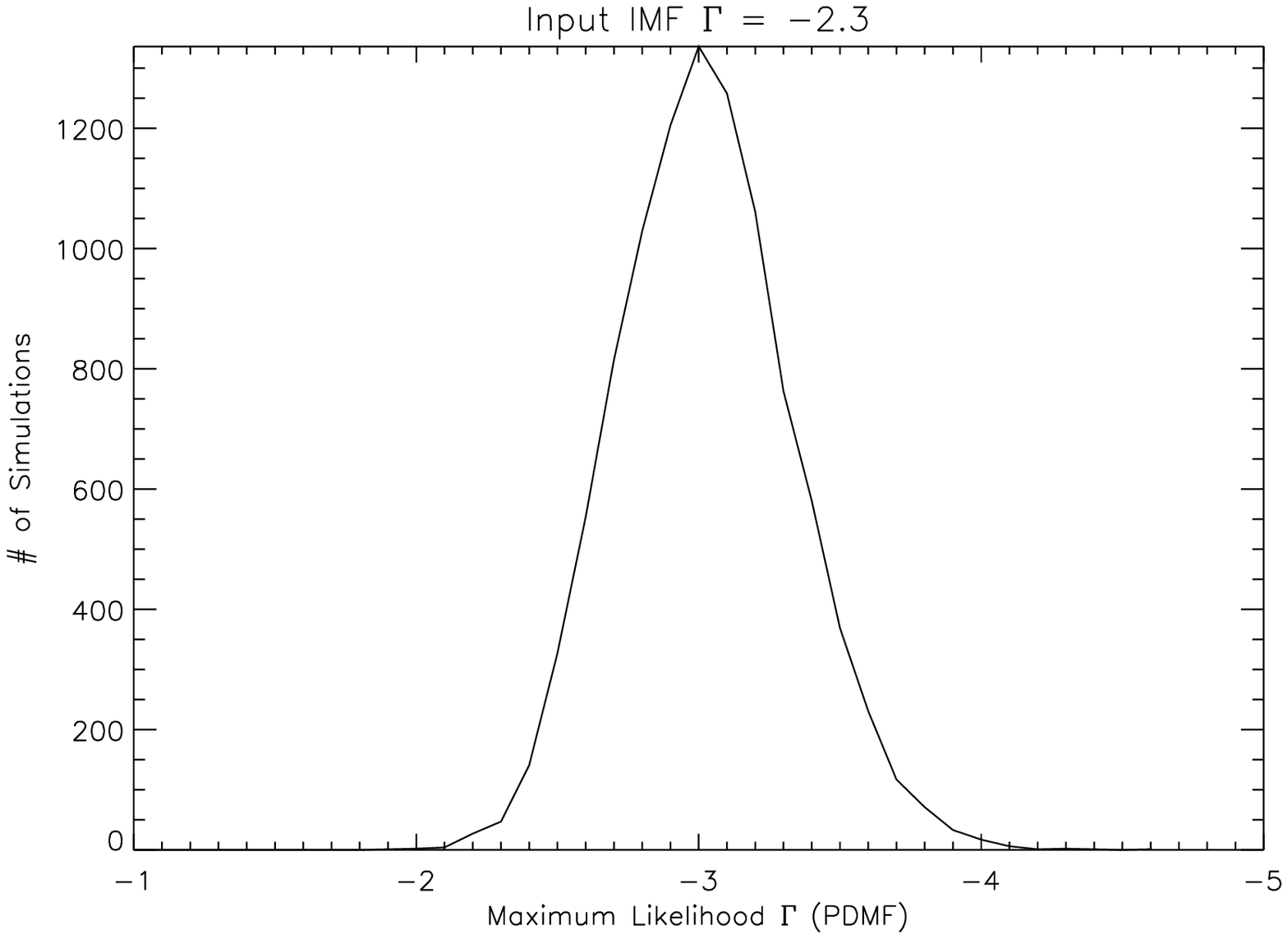}
	\caption{The distribution of PDMF slopes for an input IMF slope of $\Gamma_{\rm IMF}$ = 2.3.  Notice that these distributions peak in the range of $\Gamma$=2.8 to 3.1.}
	\label{pdmf}
	\end{center}
\end{figure*}

With an estimate of the PDMF slope, we now investigate the IMF of the SMC field.
To find the intrinsic field IMF, we employ a simple Monte Carlo simulation to create a theoretical stellar field population assuming a continuous, fixed star formation rate.  In our Monte Carlo models, we generate an artificial field population with an IMF given by $n(m)\ dm \propto m^{-(\Gamma_{\rm IMF}+1)} dm$, where we vary $\Gamma_{\rm IMF}$ between 1.0 to 4.0 in steps of 0.1.  We set $\mlo = 20M_\odot$ and $\mup = 150M_\odot$ as fixed parameters for each simulation and assign each star a random age from 0 to $\sim 10^7$ yrs (the lifetime of a 20M$_\odot$ star) to simulate continuous star formation.  
Since stellar lifetime is inversely correlated with stellar mass, we eliminate stars $> 20 M_\odot$ with assigned ages greater than their expected lifetimes.
For each value of  $\Gamma_{\rm IMF}$, we generate $10^4$ artificial field populations.  For each artificial population, we include only the first 130 stars, to provide an accurate comparison with the RIOTS4 sample.  With the $10^4$ artificial populations, we compare the resultant distribution of PDMF slopes from each model to the observed PDMF slope from the RIOTS4 sample.  We find that an input IMF of $\Gamma_{\rm IMF}$ = 2.3 generates a PDMF distribution that most closely matches the the observed PDMF values from the nonlinear least squares fit and the maximum likelihood method.  Figure \ref{pdmf} shows this distribution of PDMF slopes for $\Gamma_{\rm IMF}$ = 2.3 in both least squares fitting and maximum likelihood methods.  We can use the distribution in Figure \ref{pdmf} to estimate the error in our IMF slope using this method.  Assuming a Gaussian distribution, the 1-$\sigma$ error corresponds to a difference of $\pm 0.4$ in the slope.  Thus, we arrive at our final estimate of $\Gamma_{\rm IMF}$ = 2.3 $\pm 0.4$.  This slope is much steeper than the canonical Salpeter slope of $\Gamma_{\rm IMF}$ = 1.35.

In addition to the slope of the IMF, the upper mass limit is also of interest.  However, we argue here that our sample size is insufficient to probe this value, given the steep PDMF.  In the limit $\mup \to \infty$, the normalization condition for the mass function takes the form
\begin{equation}
\alpha = N \Gamma_{\rm PDMF} \mlo^{\Gamma_{\rm PDMF}},
\end{equation}
where we have normalized the distribution for $N = 130$ stars in the sample.  With this normalization, we can find the largest stellar mass such that we expect the sample to contain at least one star.  
This mass scale is given by
\begin{equation}
\label{mhieq}
\mmax = \mlo N^{1/\Gamma_{\rm PDMF}}
\end{equation}
where $\mlo = 20M_\odot$.  From equation (\ref{mhieq}), we expect to find no stars above $80M_\odot$ or $100M_\odot$ for $\Gamma_{\rm PDMF} = $ 3.5 or 3.0, respectively.  These values are comparable to our observed highest mass star of $75_\odot$, yet well below the putative upper mass limit of $\sim ~150_\odot$ found in Milky Way and LMC clusters (Oey \& Clarke 2005; Koen 2006).  As a result, due to the steepness of the mass function and the relatively small sample size, we find that the entire SMC field population is not large enough to contain stars larger than about $\sim 100 M_\odot$ with high probability.  In turn, this limitation implies that this sample does not constrain the upper mass limit for stars in the sparse field environment.  Larger populations in other galaxies should be studied to probe this important issue.

\subsection{Field IMF from $7-20M_\odot$}
\label{ogleimfsec}

To probe the stellar IMF of the SMC field below the detection limits of RIOTS4, we utilize $BV$ photometry of the SMC bar from Phase II of the Optical Gravitational Lensing Experiment (OGLE; Udalski et al. 1998).  Depending on the stellar density, OGLE photometry is $\sim 75-90\%$ complete to $B \sim 20$ and $V \sim 20.5$ (Udalski et al. 1998).  Assuming a distance modulus of 18.9 for the SMC (Harries et al. 2003), these magnitudes correspond to a $3M_\odot$ star along the ZAMS.  Therefore, we adopt $3M_\odot$ as the lower mass cutoff for our field star target selection from OGLE photometry.  The target selection process is similar to that of the RIOTS4 survey; however, in this case our initial selection criteria include all stars above the $3M_\odot$ Geneva evolutionary track in absolute magnitude, $M_V$, vs. $B-V$ color (Girardi et al. 2002).  In addition, we include stars that fall blueward of the main sequence evolutionary tracks with $M_V \leq$ 1, which are likely main sequence stars $\geq 3M_\odot$.  Prior to these selections, all stars are extinction-corrected using the SMC extinction maps from Zaritsky et al. (2002).  With this sample, we run a new iteration of the friends-of-friends code, just as in the target selection for the RIOTS4 survey.  Here, the relatively high density of the SMC bar and inclusion of lower mass stars results in a higher stellar surface density than in the RIOTS4 sample.  Thus, when we maximize the total number of clusters to find the clustering length (18 pc), it is lower than in the RIOTS4 iteration of this code.  Therefore, in the OGLE sample, field stars are defined as stars located at least 18 pc away from their nearest neighbor.

\begin{figure*}
	\begin{center}
	\includegraphics[scale=.65,angle=0]{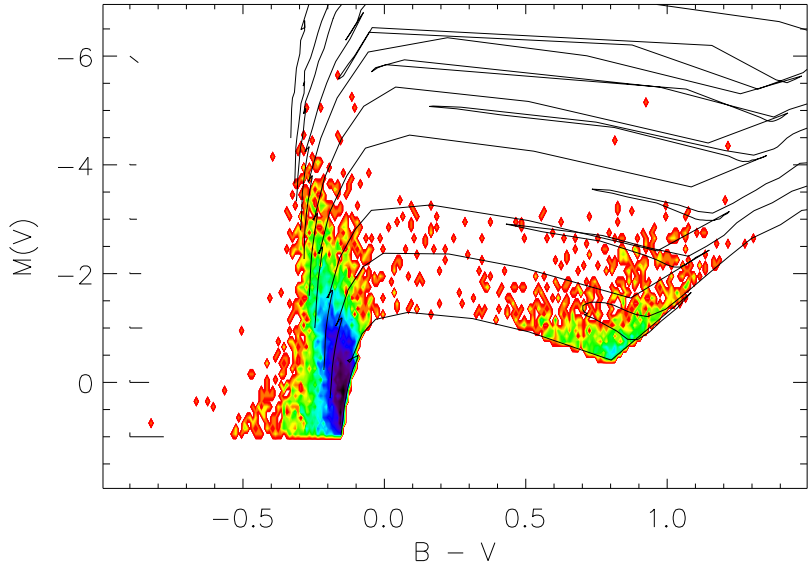}
	\caption{Contour plot depicting a color magnitude diagram of all stars in the SMC included in our initial selection criteria from OGLE $BV$ photometry.  The contours for $10^0$, $10^1$, $10^2$, and $10^3$ stars per bin begin at red, green, blue, and black, respectively.  The Geneva evolutionary tracks (Girardi et al. 2002) range from $3-40M_\odot$.}
	\label{bvcontour}
	\end{center}
\end{figure*}

\begin{figure*}
	\begin{center}
	\includegraphics[scale=.65,angle=0]{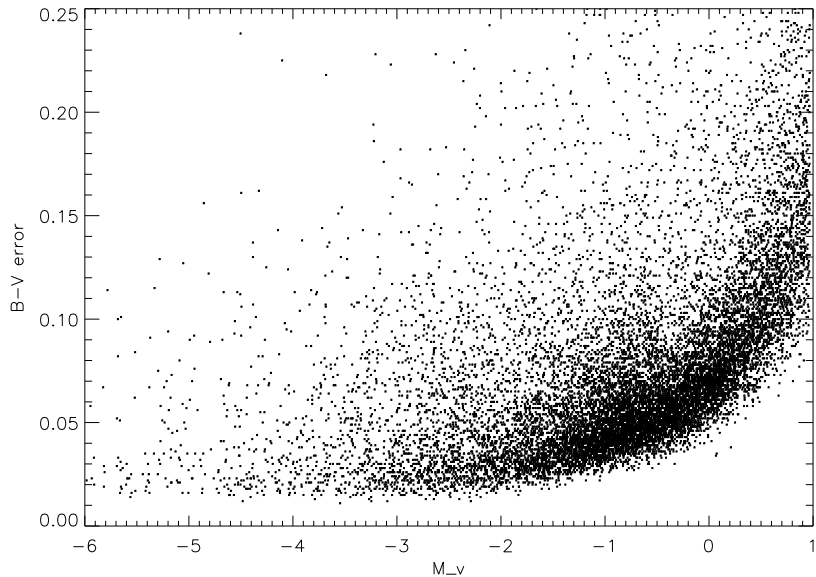}
	\caption{The distribution of errors in $B-V$ vs. $M_V$.}
	\label{bverr}
	\end{center}
\end{figure*}

Figure \ref{bvcontour} shows a color-magnitude contour plot of all stars included in the friends-of-friends algorithm.  Stars shown in Figure \ref{bvcontour} are from a single 14.2'' x 52'' OGLE field and represent $\sim 10\%$ of the full OGLE sample.  Stellar density contours of $10^0$, $10^1$, $10^2$, and $10^3$ stars per bin begin at the colors red, green, blue, and black, respectively.  We also plot evolutionary tracks between $3M_\odot$ and $40M_\odot$ from Girardi et al. (2002), which are calculated from Charbonnel et al. (1993).  The horizontal and vertical lines at $B-V$ = 0.9 depict the mean errors in $B-V$ and $M_V$, respectively, as they change with magnitude.  The line in the upper left indicates the mean error in the extinction measurement.  We also plot the distribution of $B-V$ errors vs. $M_V$ in Figure \ref{bverr}.  From Figure \ref{bvcontour}, the distribution of stars blueward of the ZAMS is consistent with a population of main sequence stars that are displaced due to photometric errors.

\begin{figure*}
	\begin{center}
	\includegraphics[scale=.65,angle=0]{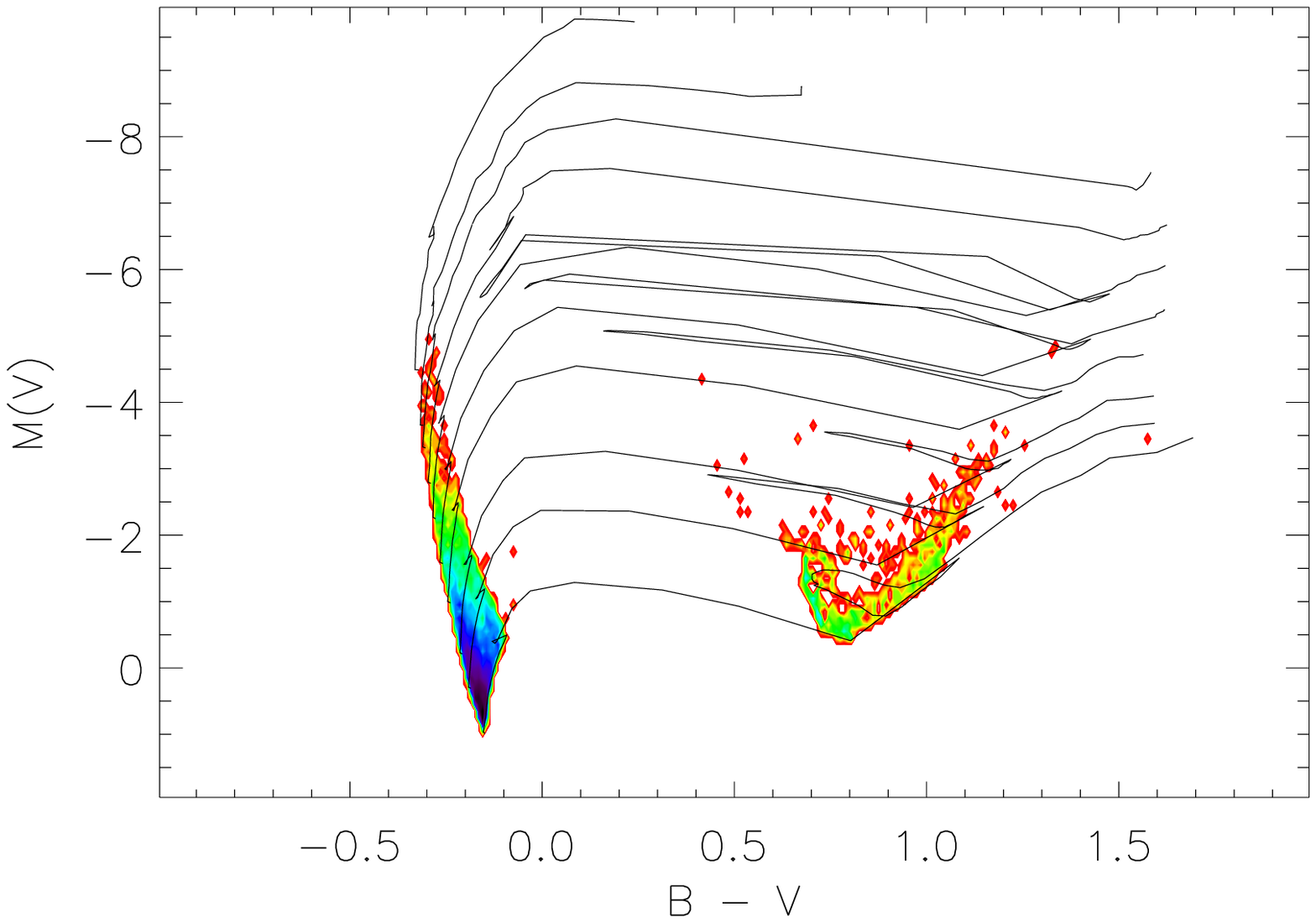}
	\includegraphics[scale=.65,angle=0]{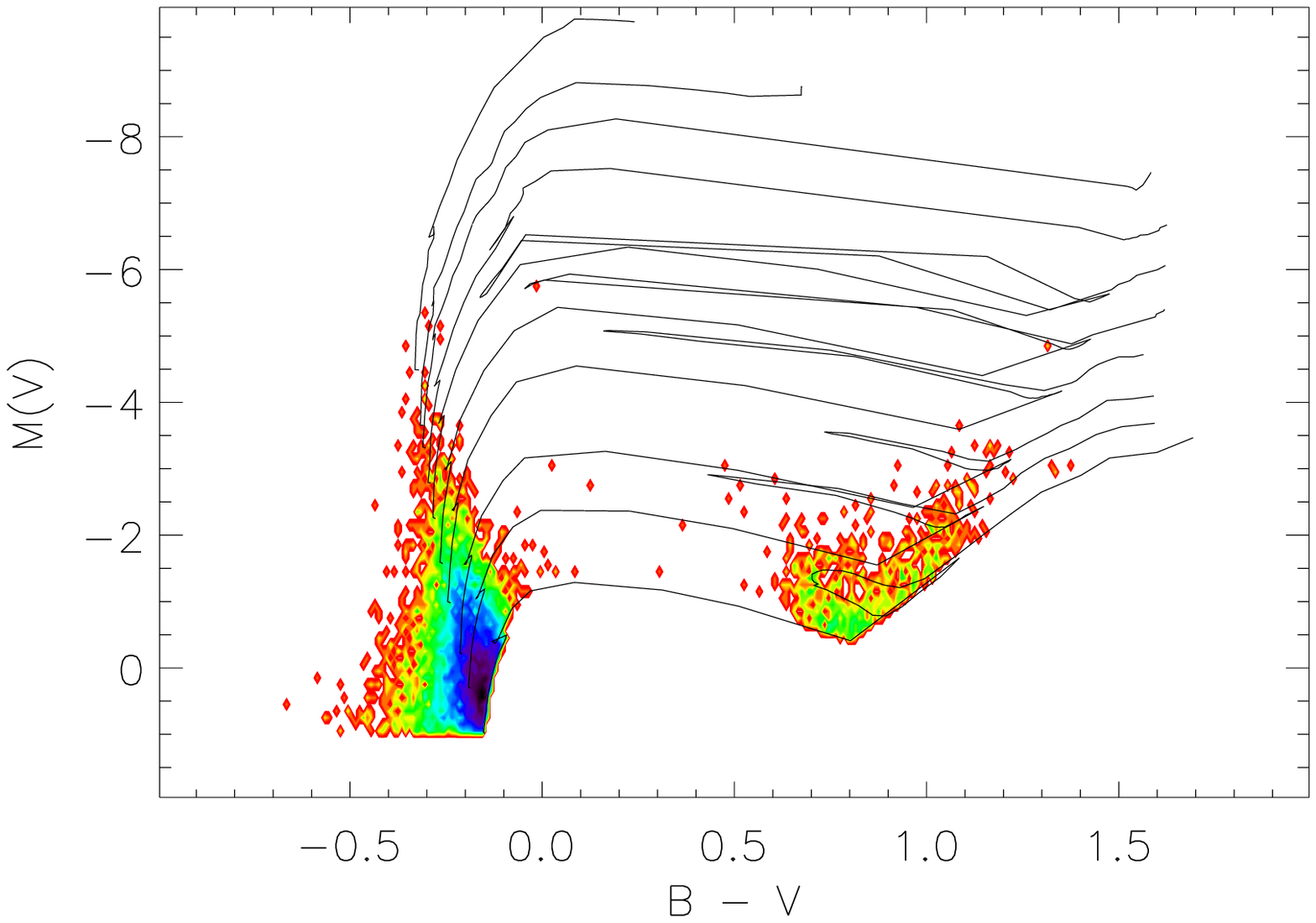}
	\caption{Axes, contours, and evolutionary tracks as in Figure \ref{bvcontour}.  The top panel depicts an artificial stellar population generated from the evolutionary tracks.  The bottom panel depicts the same population, with $B$, $V$, and extinction errors from the OGLE survey.}
	\label{bvmonte}
	\end{center}
\end{figure*}

To examine the effects of photometric errors on the OGLE CMD, we generate an artificial stellar population following the Girardi et al. (2002) evolutionary tracks and then simulate their distribution in $M_V$ vs. $B-V$ based on the uncertainties in OGLE photometry and SMC extinction values.  We create this artificial stellar population from a standard Salpeter IMF, given by $n(m)\ dm \propto m^{-(\Gamma_{\rm IMF}+1)} dm$, where $\Gamma_{\rm IMF}$= 1.35, $\mlo=2M_\odot$, and $\mup=120M_\odot$.  We again operate under the condition of continuous star formation, thus assign each star a random age from 0 to $\sim 10^9$ yrs (the lifetime of a $2M_\odot$ star).  With given mass and age, we estimate $M_V$ and $B-V$ for each star by performing a linear interpolation between the Geneva model isochrones (Girardi et al. 2002).  The top panel of Figure \ref{bvmonte} depicts this artificial population above $3M_\odot$ on a color-magnitude contour plot.  Using this simulated photometry, we model an observation of this hypothetical population.  To do this, we assign the $B$, $V$, and extinction $errors$ of a real OGLE star to each of our hypothetical stars.  We ensure that the difference in both $B$ and $V$ magnitudes between our hypothetical star and its real OGLE match is $< 0.1$.  We degrade the hypothetical photometry based on these errors by selecting a random value from a Gaussian distribution with 1-$\sigma$ given by the observed uncertainty for each error term individually.
We stress that we are not modeling the photometry of these stars, but simply applying the observed OGLE errors to our model population.

The bottom panel of Figure \ref{bvmonte} is a color-magnitude contour plot of our artificial population after applying the OGLE observational errors.  This shows how the OGLE observational errors affect a hypothetical stellar population with a Salpeter IMF.  This plot can be directly compared to the actual OGLE observations in Figure \ref{bvcontour}.  We note the generally good agreement between our artificial population and the OGLE observations as a whole.  Since we took no effort to model the actual OGLE stellar population, and only modeled the observed OGLE errors, this comparison confirms that the population blueward of the ZAMS is due to the observational errors and therefore, contains important information about the stellar IMF.  However, there are three noticeable discrepancies between these plots; the redward extension of the main sequence in the OGLE data, the difference in shape of the giant populations, and the quantity of stars between the main sequence and giant populations.  These discrepancies are not due to the errors, but rather represent a real difference between our model population and the OGLE observations.  The main sequence offset is similar to that observed in the RIOTS4 spectroscopic data (Figure \ref{photcmd}), which points to this issue arising from the stellar evolutionary models, rather than our artificial star models.  The issues between the main sequence and the giant population can be partially explained due to Galactic contamination.  We investigate this contamination using the Bescanon model to estimate the different Galactic stellar populations in the direction of the SMC (Robin et al. 2003).  We find that nearly all Galactic stars have $B-V > 0.4$, with the highest density of contamination from $B-V \sim 0.5-0.7$.  This range of $B-V$ values represents the largest deviation of the giant population between the artificial model and the OGLE data.  The number of stars observed in OGLE between $B-V \sim 0.4-0.7$ is within a factor of two of the expected Galactic contribution.

We now apply our OGLE error models not to an artificial population, but to the observed SMC field population.  Using the sample of field stars identified by our friends-of-friends code, we measure the IMF of the field with a statistical approach rather than doing a basic star count per mass bin.  This unique method is advantageous for two reasons: (1) we want to incorporate stars blueward of the ZAMS into our measurement of the IMF, and (2) the size of the color-magnitude parameter space coupled with the observational errors leads to very inaccurate mass estimates (e.g., Massey 2011).  Thus, our method should be more accurate and include more stars than the star count per mass bin method.  For each OGLE star, we generate $10^4$ unique realizations of its $B$, $V$, and extinction values by selecting a random value on a Gaussian distribution centered on the observed values, where the 1-$\sigma$ value of the distribution is given by the observational error for each measurement.
With these $10^4$ realizations, we count how many lie within each mass bin of the Geneva evolutionary tracks.  From this count, we assign a fractional probability for the star's existence in each mass bin, which is calculated by the the number of realizations in that mass bin divided by the total realizations found within all mass bins.  Any realizations that fall outside the Geneva evolutionary tracks are ignored as unphysical realizations and do not count towards this analysis.  This ensures that each star is weighted equally for the IMF, by having a total probability over all mass bins equal to one.  If we did not exclude unphysical realizations, then a star directly on the ZAMS would only have a total probably of $\sim 0.5$, since about half its realizations would exist blueward of the main sequence, while a star near the turnoff of the main sequence would have nearly all its realizations counted.  Thus, we ensure that stars blueward of the main sequence are counted equally to stars near on the main sequence. 

Before completing the IMF measurement, we first re-evaluate the issue of completeness for the IMF sample.  We have demonstrated that stars scattering between mass bins is important, and therefore, we want our data to be complete below the mass bins we consider in our IMF measurement.  In addition, some OGLE fields are only 75\% complete for a ZAMS star of $3M_\odot$, assuming no extinction is present.  Thus, to minimize completeness problems, we adopt the lower mass limit for this IMF measurement to be $7M_\odot$.  Even for OGLE fields with the highest stellar density, completeness for $7M_\odot$ stars on the ZAMS is $>95\%$ with $A_V$=0 or $>90\%$ with $A_V$=1.  

\begin{figure*}
	\begin{center}
	\includegraphics[scale=.65,angle=0]{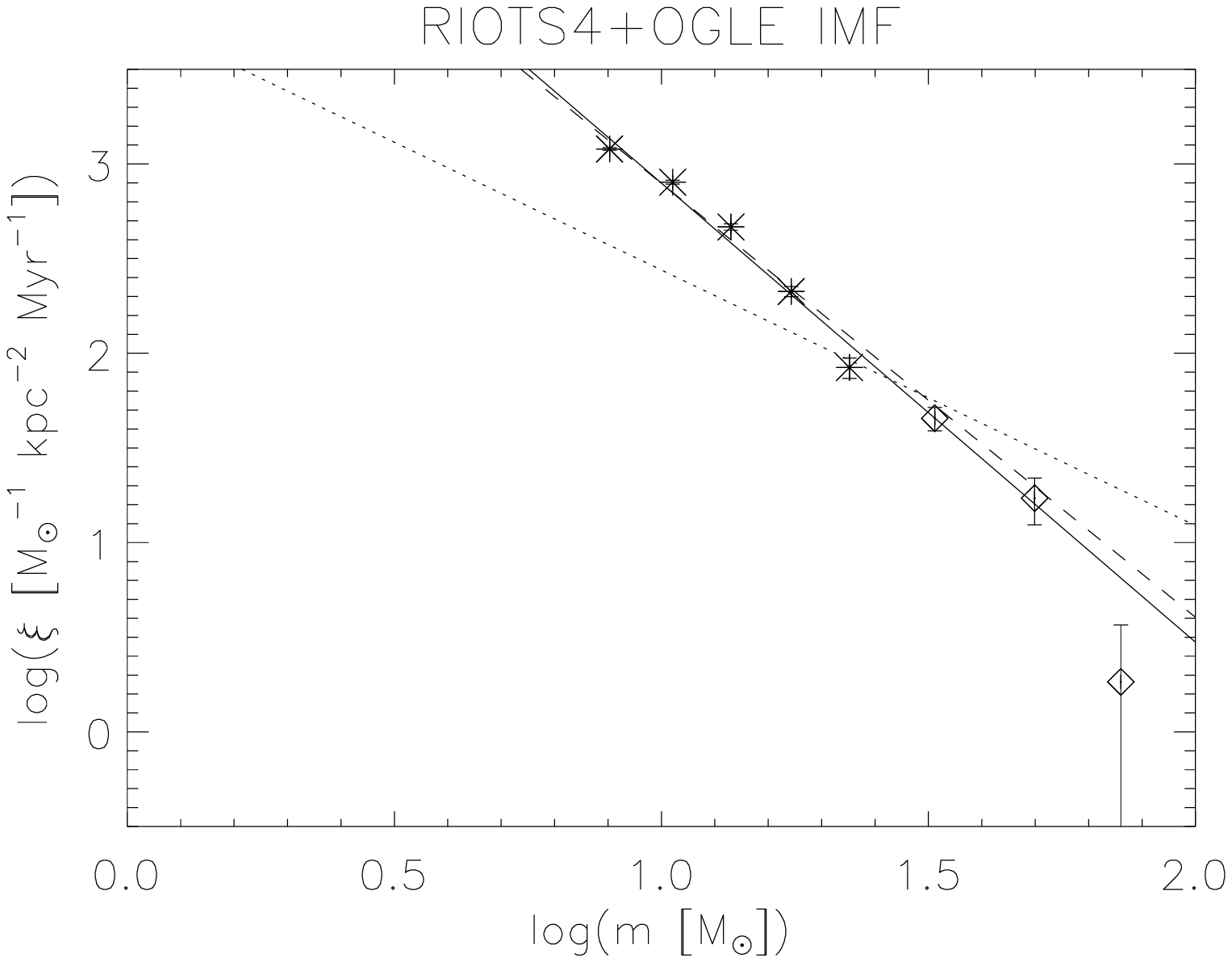}
	\caption{The IMF of the SMC field star population as derived from the OGLE $BV$ photometry (dashed line) and combined OGLE plus RIOTS4 data (solid line).  The slope of the dashed line is $\Gamma_{\rm IMF}$ = 2.3 and solid line is $\Gamma_{\rm IMF}$ = 2.4.  For reference, a Salpeter slope $\Gamma_{\rm IMF}$ = 1.35 is plotted as a dotted line.}
	\label{ogleriots4imf}
	\end{center}
\end{figure*}

With our new lower mass limit of $7M_\odot$, we plot the IMF from our statistical analysis in Figure \ref{ogleriots4imf}.  Here, we follow the formalism of Scalo (1986), where $\log \xi$ represents the mass function in units of stars born per unit mass($M_\odot$) per unit area (kpc$^2$) per unit time (Myr).  In Figure \ref{ogleriots4imf}, OGLE data are plotted as asterisks, with error bars given by the poisson uncertainties. The dashed line is a linear fit to these points, which is weighted by the Poisson errors.  This fit yields $\Gamma_{\rm IMF}$ = 2.3 $\pm 0.6$.  The dotted line shows a Salpeter slope of $\Gamma_{\rm IMF}$ = 1.35.  The most striking feature of Figure \ref{ogleriots4imf} is that the field IMF from $7-25 M_\odot$ appears to gradually transition from a steep high mass slope to a Salpeter-like slope at lowest two mass bins.  This turnover in the mass function may be an indication that different processes are driving the low and high mass IMF slopes of the field.

\subsection{Combined Field IMF}

With the combination of OGLE photometry and RIOTS4 spectroscopy, we are now able to fully characterize the IMF of the SMC field population above $7M_\odot$.  In Figure \ref{ogleriots4imf}, we also include a binned mass function from the RIOTS4 data, plotted as diamonds.  With the superior accuracy of masses derived by spectral types, we simply tally the number of stars in each mass bin to construct the IMF above $20M_\odot$ for the RIOTS4 data.  Due to the different methods used to identify the OGLE and RIOTS4 field samples, they must be normalized to one another.  Therefore, we use the $20M_\odot$ to $25M_\odot$ mass bin, which is common to both these two data sets, to normalize the RIOTS4 star count to the OGLE star count.  We apply this normalization to each mass bin in the RIOTS4 data to construct the full IMF shown in Figure  \ref{ogleriots4imf}.  Here, we see that nearly the entire mass range can be well described by a single power law with slope of $\Gamma_{\rm IMF}$ = 2.4 $\pm 0.4$ (solid line), in full agreement with our slope derived from the cumulative distribution function of the RIOTS4 masses ($\Gamma_{\rm IMF}$ = 2.3 $\pm 0.4$).  Only at the lowest bins does a turnover in the power law begin to appear.  We emphasize that this turnover is not an effect of incompleteness, since the OGLE data is $>90\%$ complete for a $7M_\odot$ ZAMS star with $A_V = 1$, whereas the typical extinction towards the SMC is $\sim$ 0.5 mag (Zaritsky et al. 2002).

\section{Discussion}
 
\subsection{Effects of Binary Star Systems}
\label{binarysec}

Recent studies indicate that the massive star binary fraction is quite high, with observations of open clusters (e.g. Sana et al. 2008, 2009, 2011) and massive clusters (Kiminki \& Kobulnicky 2012) indicating lower limits of 60\% and 70\%, respectively.  Our own analysis of the field binary fraction using $\sim 10$ epochs of observations for 30 RIOTS4 stars reveals a binary fraction $>50$\% (Lamb et al. in prep).  However, the fraction of RIOTS4 spectra that exhibit clear indications of binarity is small.   Thus, a large fraction of undetected binaries is a concern for our IMF measurements.  If a binary system is treated as a single star, the excess flux will result in an overestimate of the mass of the primary.  Additionally,  if the secondary star is $>20M_\odot$, then its absence from the IMF will further bias the star count in favor of the higher mass bins.  To quantify the magnitude of this effect, we design a simple Monte Carlo code to examine the ramifications of undetected binaries on our observed PDMF in RIOTS4.

In this analysis, we assume that each object in the RIOTS4 survey is in an undetected binary system.  The primary stars in these binary systems are assumed to have the OGLE4 derived spectral types, with a possible mass range from $20M_\odot$ to $75M_\odot$.  We randomly assign each binary system a mass ratio, $q = m_2/m_1$, which is uniformly distributed from 0.01 to 1.  This uniform distribution in $q$ is motivated by recent observational studies of the binary mass ratios in open and massive clusters (see e.g. Sana \& Evans 2011; Kiminki \& Kobulnicky 2012).  In comparison with a Salpeter distribution of secondary masses, this uniform distribution in $q$ will have a higher fraction of massive secondaries and more strongly affect the IMF results.  With these mass ratios, we use a simple power law mass-luminosity relationship to split the observed light into two separate binary components.   We recalculate the magnitude of the primary star using
\begin{equation}
m_{\rm bol,1} = m_{\rm bol} - 2.5 \log \Big{[}\frac {1} {1+q^\delta}\Big{]} \ ,
\end{equation}
where $m_{\rm bol}$ is the derived bolometric magnitude of the binary system, $m_{\rm bol,1}$ is the bolometric magnitude of the primary star, $q$ is the mass ratio of the binary system, and $\delta$ is the power law of the mass-luminosity relationship, given by $L \propto m^\delta$.
With $m_{\rm bol,1}$ and $T_{\rm eff}$, we derive the mass of the primary as in \S \ref{HRDsection} under the assumption that the observed stellar spectrum, and thus our adopted $T_{\rm eff}$, accurately reflect the physical properties of the primary star.  With these derived primary masses and secondary masses given directly by $q$, we recreate the RIOTS4 PDMF with a 100\% undetected binary fraction and measure its slope accordingly.  We perform this analysis $10^3$ times for each of three different mass-luminosity relationships, given by $\delta = 1$, $\delta = 2$, and $\delta = 3$.  For simplicity, we opt to include only single power law models, rather than a broken power law for the mass-luminosity relationship.  These different values of $\delta$ encompass the range of values expected at both high stellar masses $\delta \sim 1$ and lower stellar masses $\delta = 3$ and represent the cases of maximal and minimal impact on the IMF, respectively. 

 \begin{figure*}
	\begin{center}
	\includegraphics[scale=.65,angle=0]{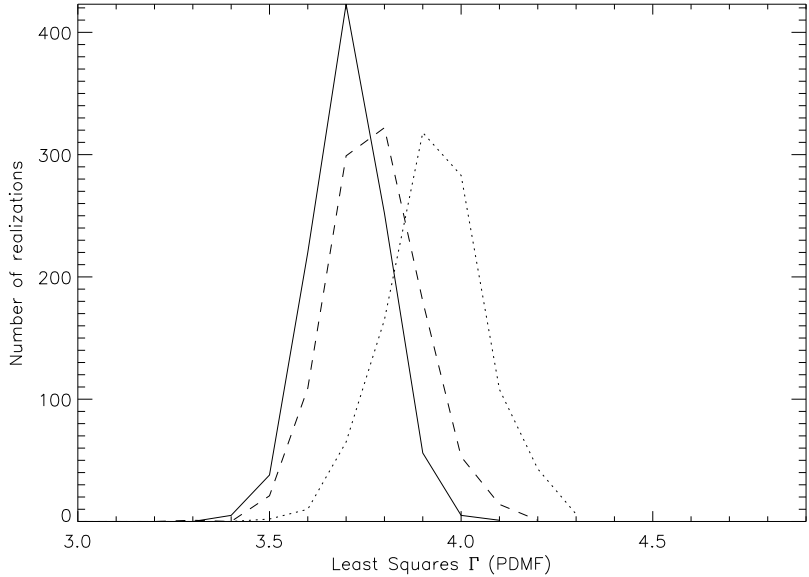}
	\caption{The distribution of PDMF slopes for RIOTS4 if all stars are in undetected binary systems.  The solid, dashed, and dotted lines represents simulations with $\delta = 3$, 2, and 1, respectively.  We display slopes from the linear least squares fit for this analysis, for which our derived RIOTS4 PDMF slope is $\Gamma_{\rm PDMF} = 3.5$. }
	\label{binaryslopes}
	\end{center}
\end{figure*}

The distribution of linear least squares PDMF slopes for these Monte Carlo simulations can be found in Figure \ref{binaryslopes}.  Simulations with $\delta = 1$, 2, and 3 are plotted with dotted, dashed, and solid lines, respectively.  These distributions can be directly compared with our measured RIOTS4 linear least squares PDMF slope of $\Gamma_{\rm PDMF} = 3.5$.  Figure \ref{binaryslopes} shows that undetected binaries will only cause a steepening of the RIOTS4 PDMF slope.  The degree of steepening ranges from 0.2 to 0.4 and depends weakly on the power law slope of the mass-luminosity relationship.  Finally, we perform one additional simulation of the extreme case where all RIOTS4 stars are in undetected equal mass binaries, which results in $\Gamma_{\rm PDMF} = 4.5$.  Thus, we clearly demonstrate that undetected binary systems will only serve to steepen the PDMF and therefore, IMF of the field.  Thus, undetected binaries cannot be the source of the steep high mass field IMF.

\subsection{Star Formation History}

\begin{figure*}
	\begin{center}
	\includegraphics[scale=.65,angle=0]{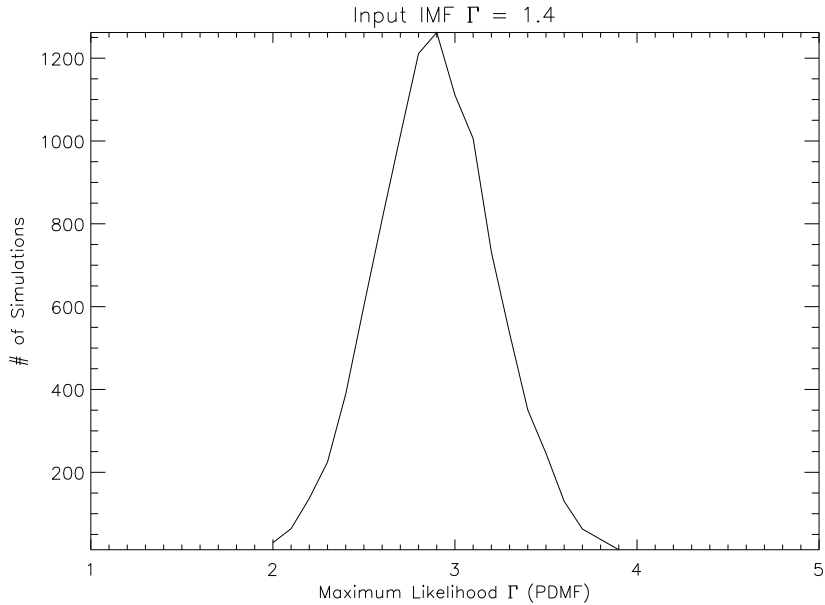}
	\caption{The distribution of PDMF slopes for the best burst model.  As in Figure \ref{pdmf} these distributions agree with the observed RIOTS4 PDMF slope of $\Gamma = 2.8 - 3.1$.}
	\label{pdmfsim}
	\end{center}
\end{figure*}

 One critical assumption made thus far in this work is that the star formation rate in the SMC has remained constant for the last $\sim$ 10 Myr.  Here, we investigate the possibility that a recent burst of star formation could be responsible for the steep field IMF.  To perform this analysis, we proceed as in Section \ref{riotsimfsec} by generating a theoretical population and measuring its PDMF for the purposes of comparing it with the observed RIOTS4 PDMF.  For each model, in addition to a continuous star formation rate, we add a burst of star formation.  However, here, we always use a standard Salpeter IMF to generate both the continuous and bursting stellar populations.  With these models, we investigate how the time since the burst, duration of the burst, and the star formation rate of the burst affect the output PDMF of the model.  We restrict our models to bursts beginning within the last 10 Myr, which is approximately the lifetime of a $20M_\odot$ star.  For each set of model parameters, we generate $10^4$ theoretical populations and as before, we plot the distribution of PDMF slopes for each model.  Covering each variable in turn, we find that the higher the star formation rate of the burst, the more it steepens the PDMF.  The duration of the burst, however, exhibits more subtle behavior.  A burst too long in duration tends to mimic the PDMF of continuous star formation.  Likewise, a burst that is too short also does not significantly alter the PDMF.  In most cases, the maximum effect on the PDMF occurs for a burst duration of $\sim$ 5 Myr.  Finally, we find that bursts ending $<$ 3 Myr ago generate too many high mass stars, making these models incompatible with our observed slope and highest mass star.

In addition to tracking the distribution of PDMF slopes, we also investigate how the star formation history affects the shape of the PDMF and IMF.  Since the shape of individual models can differ significantly from one population to the next, we generate a ranked order of masses for each iteration of the model and take the average value at each rank to create the model PDMF.  For example, we take the most massive star from each iteration for a given model and average these values to get the most massive star of the model PDMF.  This process continues for each of the N stars in the burst models, where N is the number of stars $> 20M_\odot$ from the RIOTS4 survey.  Since there is no guarantee that we will form a given number stars in a given model, we always take the first 130 stars $> 20 M_\odot$ to ensure a direct comparison to our RIOTS4 sample.  We also extend this ranked average of stars down to $7 M_\odot$ so we can make a direct comparison of the averaged IMF of these models with the combined RIOTS4 and OGLE IMF.

\begin{figure*}
	\begin{center}
	\includegraphics[scale=.65,angle=0]{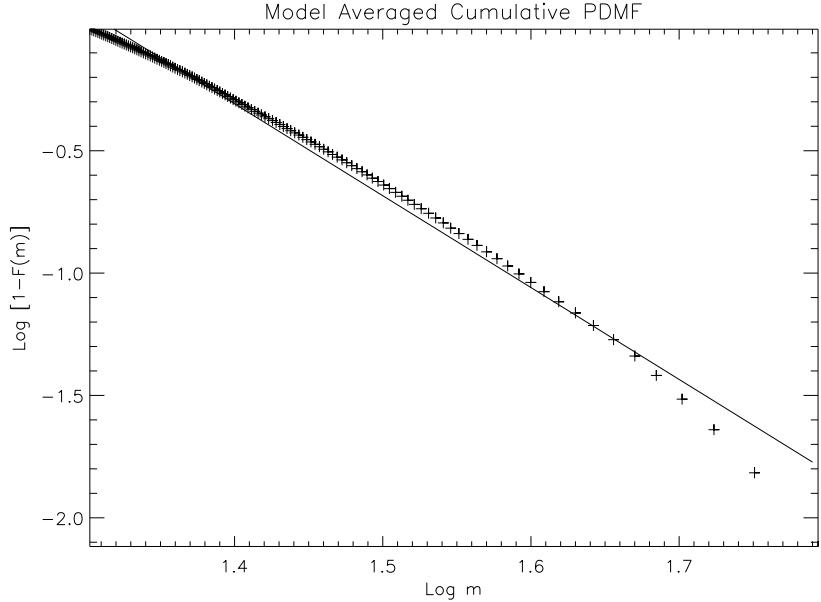}
	\caption{The averaged PDMF of the best burst model, plotted as in Figure \ref{koen}.  The solid line is a linear least squares fit to this model with slope $\Gamma_{\rm PDMF}$ = 3.7}
	\label{pdmfdistsim}
	\end{center}
\end{figure*}

\begin{figure*}
	\begin{center}
	\includegraphics[scale=.65,angle=0]{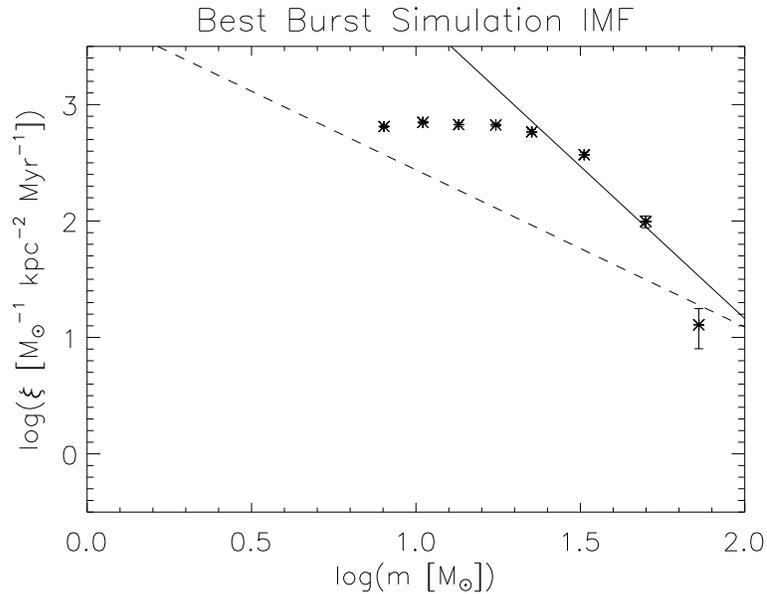}
	\caption{The averaged IMF of the best burst model, plotted as in Figure \ref{ogleriots4imf}.  The solid line shows a linear least squares fit for data above $20M_\odot$, which has a slope of $\Gamma_{\rm IMF}$ = 2.6.  The dashed line shows a Salpeter IMF as reference.  Notice the steep turnover below $20M_\odot$ in comparison with Figure \ref{ogleriots4imf}.}
	\label{burstsimimf}
	\end{center}
\end{figure*}

The model that most closely matches the observed PDMF is a burst of star formation that begins 8 Myr ago, lasts for 5 Myr, and does not include any continuous star formation component along with the burst.  Figure \ref{pdmfsim} shows the distribution of PDMF slopes for this best burst model, which are in agreement with the $\Gamma = 2.8 - 3.1$ observed for RIOTS4.  Figure \ref{pdmfdistsim} depicts the averaged cumulative PDMF for the best burst model, which can be directly compared with the observed PDMF from RIOTS4 (Figure \ref{koen}).  Similarly, Figure \ref{burstsimimf} shows the averaged IMF of the best burst model and can be compared with the RIOTS4 and OGLE combined IMF.  To compare the averaged PDMF distribution with the observed RIOTS4 PDMF distribution, we use a Kolmogorov-Smirnov (KS) test and find a 90\% likelihood that these populations were drawn from the same parent population.  However, this agreement does not extend below $20 M_\odot$.   As seen in Figure \ref{burstsimimf}, the averaged IMF of the best burst model turns over much more rapidly below $20 M_\odot$ than the combined RIOTS4 and OGLE IMF.  A KS test comparing these two populations still finds a 68\% likelihood they were drawn from the same parent population, but with the small number of data points, this percentage may be an overestimate.  

Therefore, we find some evidence that for a narrow range of burst parameters, it is possible that the observed RIOTS4 PDMF of the SMC field can be reconciled with a Salpeter IMF.  However, with inclusion of the OGLE data, the shape of the mass function does not represent a smooth power law as seen in the combined RIOTS4 and OGLE IMF.  Thus, it is unlikely that the observed PDMF is due solely to a recent burst of star formation.  

Another possible explanation for the steep IMF is a decline in the star formation rate.  If such a decline occurs over the lifetime of the least massive stars included in our IMF, then our derived IMF will be relatively steeper than the actual IMF.  Elmegreen \& Scalo (2006) calculate the necessary decline in star formation rate such that a Salpeter IMF will appear to have $\Gamma_{\rm IMF}$= 2.0, 3.0, and 4.0.  We focus on the cases of $\Gamma_{\rm IMF}$= 2.0 and 3.0, which are reasonable estimates for the lower and upper bounds on our derived IMF slope above $7M_\odot$.  To derive $\Gamma_{\rm IMF}$= 2.0 or 3.0, the star formation rate must decrease by a factor of 2.5 or 20, respectively, over the last 50 Myrs, the lifetime of a $7M_\odot$ star.   Thus,  a decline in star formation rate by approximately an order of magnitude over the last 50 Myrs would be sufficient to explain the steep IMF of $\Gamma_{\rm IMF}$= 2.4.  However, studies of the recent star formation history of the SMC point to a largely uniform rate of star formation over the past $\sim 50$ Myrs (e.g. Harris \& Zaritsky 2004; Chiosi et al. 2006; Indu \& Subramaniam 2011).  Therefore, it is also unlikely that the steep IMF is a result of a recent decline in the star formation rate.

\subsection{Runaways}
Runaways are a well-known component of the massive star field population; however, their contribution to the field population remains uncertain.  While the fraction of O star runaways is found to be higher than for B stars (eg. Blaauw 1961; Stone 1991), the total fractional contribution of runaways to the field population remains controversial (see \S 1).  Moreover, there are other complicating factors that affect the properties of the runaway population including the runaway fraction as a function of stellar mass, the method and timescale of ejection, and, tied to this, the fraction of stellar lifetime spent as a runaway.  In most ejection scenarios, either dynamically or through a supernova kick, it is the secondary object in a binary system that gets ejected.  Therefore, it may be expected that a turnover exists in the runaway fraction at a mass scale less than the maximum mass $m_{up}$.  If this turnover were to occur in the middle of the RIOTS4 observed mass range, the runaway contribution may introduce a local maximum in the PDMF.  In contrast, a direct correlation of runaway fraction with stellar mass that extends to $\mup$ would simply flatten the PDMF.  Runaways may be dynamically ejected shortly after the stars form, while the supernova ejection mechanism is delayed by the lifetime of the primary star in the binary system.  If dynamical ejections occur on a fixed timescale independent of stellar mass, then B stars will spend a greater fraction of their lives as runaways than O stars.  Conversely, if the dynamical ejection timescale inversely correlates with mass or is uniformly random over the entire lifetime of all massive stars, then the fraction of stellar lifetime spent as a runaway star may be greater for O stars than B stars.  However, the dominant ejection mechanism and the typical ejection timescale are still unknown, and therefore, the effect that runaways have on the PDMF of the field population is also unknown.

Here, we examine the observational evidence for runaways in our RIOTS4 sample by comparing stellar radial velocities with the local HI gas velocities.  Stellar radial velocities are obtained from gaussian fits to the Hydrogen Balmer series, He I, and, when applicable, He II absorption lines.  Full details will be presented in a forthcoming overview paper of the RIOTS4 survey  (Lamb et al. 2013, in prep).    
We compare the observed radial velocity for each star with the HI velocity distribution within a $1\arcmin \times 1\arcmin$ box centered at the star's position.  We identify runaway stars as those having a difference $> 30$km/s from any HI gas at this position having a brightness temperature $> 20$ K.  These conservative criteria identify ten runaway stars having mass $> 20M_\odot$.  We find that the identified runaways are preferentially higher in mass than the typical star from the RIOTS4 IMF sample; the average mass of all RIOTS4 stars with mass $> 20 M_\odot$ is $28.3M_\odot$, while the average mass of the runaway sample is $36.1M_\odot$, which includes 7 stars $>35 M_\odot$.  

\begin{figure*}
	\begin{center}
	\includegraphics[scale=.65,angle=0]{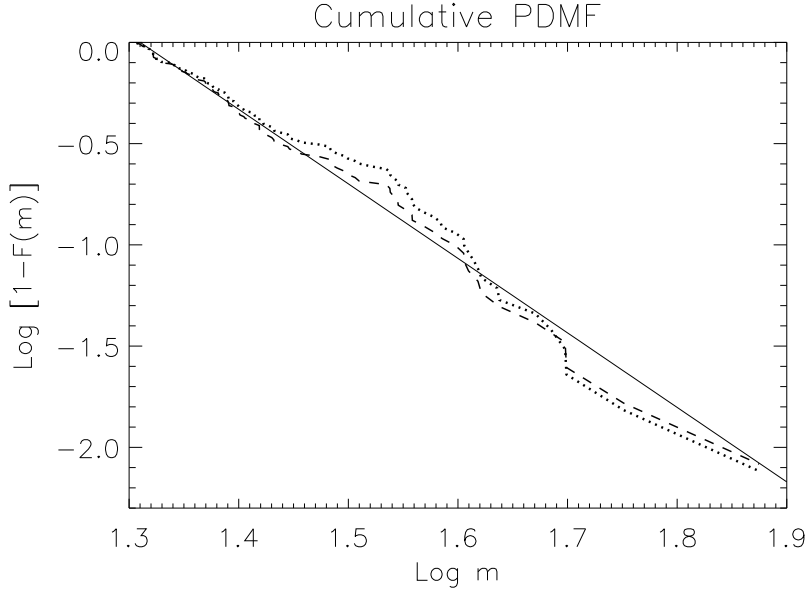}
	\caption{The PDMF of the SMC field star population with runaway stars included (dotted line) and omitted (dashed line), plotted as $\log[1-F(m)]$ versus $\log m$, where $F(m)$ is the empirical CDF.  The dotted line is the same PDMF plotted in Figure \ref{koen}.  The solid line shows the linear least squares fit to the data, with a slope of $\Gamma_{\rm PDMF}$=3.7.}
	\label{norunaways}
	\end{center}
\end{figure*}

In Figure \ref{norunaways}, we compare the RIOTS4 PDMF with runaways included (dotted line) and omitted (dashed line).  With the runaways removed, we find a steeper slope for the PDMF of $\Gamma_{\rm PDMF}$=3.7 for the linear least squares fit (solid line) and $\Gamma_{\rm PDMF}$=3.4 for both the non-linear least squares fit and the maximum likelihood method.  A visual comparison between the PDMF with runaways included and excluded reveals that the removal of the runaway stars has lowered a hump in the PDMF around $\log m \sim 1.5$.  As mentioned previously in this section, such a feature may be due to a turnover in the general trend of runaway fraction increasing with stellar mass.  Due to our detection of only line-of-sight runaways, the persistence of the hump after removing these stars could be attributed to undetected transverse runaways.  If the PDMF with runaways excluded is a closer representation of the field population that formed in situ, then following the same analysis in \S \ref{riotsimfsec} yields an intrinsic field IMF slope of $\Gamma_{\rm PDMF}$=2.8$\pm 0.5$.  Thus, we find that the removal of runaway stars from our RIOTS4 sample is consistent with a steeper in situ field IMF.

\section{Conclusions}
We conduct an extensive survey targeting a spatially complete sample of field massive stars in the Small Magellanic Cloud (SMC).  This Runaways and Isolated O Type Star Spectroscopic Survey of the SMC (RIOTS4) uses the IMACS multi-object spectrograph on the Magellan Baade telescope and the MIKE echelle spectrograph on the Magellan Clay telescope.  Targets for RIOTS4 come from Oey et al. (2004), who identify a sample of 374 candidate field massive stars using photometry to select massive stars and a friends-of-friends algorithm to ensure isolation.  A total of 284 objects yield spectra of sufficient quality to derive their effective temperatures and calculate their bolometric luminosities with photometry from Massey (2002).  These physical properties yield stellar mass estimates typically accurate to $\sim 2 M_\odot$.  The majority of stars without derived physical properties are Oe/Be stars with weak or filled-in diagnostic lines, which prevents accurate spectral classification.  

RIOTS4 is complete to $\sim 20M_\odot$ along the main sequence.  We therefore include the 130 stars with mass $\geq 20M_\odot$ in our analysis of the stellar IMF of the field.  Following the methodology of Koen (2006), we obtain a slope of $\Gamma_{\rm PDMF}$ = 2.8 to 3.5 for the present day mass function (PDMF) of the field population.  To obtain the field IMF, we use a simple Monte Carlo model to generate $10^4$ artificial field populations for a variety of input IMF slopes ranging from $\Gamma_{\rm IMF}$=1.0 to 4.0.  Assuming continuous star formation, we measure the distribution of the PDMF slopes of these different models.  The model PDMF distribution that best matched our observed PDMF slope was for an input $\Gamma_{\rm IMF}$=2.3$\pm 0.4$.  Thus, we find that the field IMF is much steeper than the canonical Salpeter slope of $\Gamma_{\rm IMF}$=1.35.  

To extend our field IMF to lower masses, we use $BV$ photometry from OGLE.  Using selection techniques similar to those for RIOTS4 targets, we identify a sample of field stars $\geq 7M_\odot$ from the OGLE survey.  We employ a statistical methodology to measure the field IMF from the OGLE data.  We randomly select values from a Gaussian distribution of the individual $B$, $V$ and extinction errors, which we use to generate $10^4$ realizations of the photometry for each OGLE field star.  We treat these realizations as a probability distribution in $M_V$ vs. $B-V$, and divide each star fractionally into different mass bins by counting the proportional contribution of realizations to each bin.  In this way, we account for stars that were displaced outside of the evolutionary models due to photometric errors.  From the OGLE data, we derive an IMF slope of $\Gamma_{\rm IMF}$=2.3$\pm 0.6$ for the mass range $7M_\odot - 20M_\odot$.  This IMF exhibits a steep slope at higher mass and gradually turns over into a slope approaching Salpeter, $\Gamma_{\rm IMF}$=1.35, at the lowest two mass bins.   We combine this OGLE IMF measurement with an IMF derived from counting the number of stars per mass bin of the RIOTS4 data to get a field IMF above $7M_\odot$.  The slope for this combined IMF is $\Gamma_{\rm IMF}$=2.4$\pm 0.4$, which again is steeper than the standard Salpeter slope (Figure \ref{ogleriots4imf}).  

One of the major potential sources of error in our IMF is the presence of a significant, undetected binary population.  To account for this possible scenario, we model the RIOTS4 population as if every star is a member of a binary system.  We select the binary mass ratio $q=m_2/m_1$ randomly from a uniform distribution from 0.01 to 1 and split the light according to a mass-luminosity relationship, $L \propto m^\delta$, where $\delta = 1$, 2, or 3.  We find that this 100\% binary fraction has a steepening effect on the observed RIOTS4 PDMF slope, in the range of 0.2-0.4, depending on the value of $\delta$, with a steeper power law having a smaller effect on the IMF.  

We investigate the possibility that a unique star formation history of the SMC combining with a Salpeter IMF could result in the observed steep PDMF.  We model a range of bursting star formation histories, by varying the burst duration, burst strength, and time elapsed since the burst occurred.  We find one `best burst' model that closely matches the $\Gamma = 2.8 - 3.1$ slope of the PDMF from RIOTS4.  This is a burst that begins 8 Myrs ago, lasts for 4.5 Myrs, and has no other star formation occurring during the previous 10 Myrs (100\% burst).  However, the `best burst' model IMF flattens dramatically below $20M_\odot$, which is a feature not seen in the OGLE IMF.  Alternatively, a steady decline in the star formation rate of the field may also steepen the observed IMF.  However, the recent star formation history of the SMC points to a continuous, rather than declining star formation rate.  We therefore conclude that a unique star formation history can not fully explain the steep SMC field star IMF.  

Finally, we identify ten runaway stars within the RIOTS4 survey, which have radial velocities at least 30 km/s from significant HI masses in the line of sight.  These runaways have an average mass of $36.1M_\odot$, which is $8.2M_\odot$ higher than the average mass of all stars in the RIOTS4 sample.  Removing these runaways from our sample steepens the observed PDMF slope from $\Gamma_{\rm PDMF} = 2.8 - 3.5$ to $\Gamma_{\rm PDMF} = 3.4 - 3.7$ and thus also steepens the intrinsic field IMF slope to $\Gamma_{\rm IMF}=2.8\pm 0.5$.  These results are consistent with the observation that runaways are more common at higher masses (eg. Stone 1991).  The runaway population also appears to correlate with a hump-like feature in the RIOTS4 PDMF around $\sim 35M_\odot$.  Such a feature could be explained by a model in which the runaways, which are relatively massive stars, are ejected by stars of even greater mass.

In summary, we employ a number of methods to measure the IMF of the SMC field using spectroscopy from the RIOTS4 survey combined with photometry from Massey (2002) and photometry from the OGLE survey.  In these analyses, we find an IMF slope of $\Gamma_{\rm IMF}$ = 2.3 - 2.4 for the high mass SMC field population, which is significantly steeper than the standard Salpeter IMF with $\Gamma_{\rm IMF}$ = 1.35.  Although not as steep, our results confirm those of previous studies of the SMC, which found a steep field IMF slope of $\Gamma \sim 3 - 4$ for stars $> 25M_\odot$ (Massey et al. 1995; Massey 2002).  In principle, the upper mass cutoff could also be different in clusters and in the sparse field environment considered herein.  In practice, however, the SMC sample size is too small to constrain the upper mass cutoff for the field population.  Nonetheless, we find a steep slope to the field IMF, and this finding could indicate that massive stars in the field do not form in exactly the same manner as those in clusters.

\acknowledgments
We thank the anonymous referee for valuable comments that strengthened the results of this work.  We are grateful to Radek Poleski for investigating issues we encountered with the OGLE survey data.  We thank Eric Pellegrini, Anne Jaskot and Jordan Zastrow for useful comments on an early draft of this manuscript.  We also thank Rupali Chandar, Oleg Gnedin, Wen-hsin Hsu, and Mario Mateo for useful discussions.  We appreciate the support and hospitality of the staff at Magellan and Las Campanas Observatory during the RIOTS4 observing runs.
This work was supported by funding from NSF grant AST-0907758.

\end{document}